\documentclass[aps,prl,floatfix,amsmath,amssymb,twocolumn,showpacs,superscriptaddress]{revtex4-2}
\usepackage{amsmath, amssymb, graphics, mathrsfs, bm, stackrel,bbold}
\usepackage{parskip}
\usepackage{subfigure}
\usepackage[usenames,dvipsnames]{xcolor}
\definecolor{Highlight}{rgb}{1,1,0.75}

\allowdisplaybreaks[1]
\usepackage{graphicx}
\usepackage[bookmarks={false}, pdfauthor={Rudro Rana Biswas}, pdftitle={Title}]{hyperref}
\hypersetup{colorlinks=true, linkcolor=BrickRed, citecolor=Violet, filecolor=OliveGreen, urlcolor=RoyalBlue, filebordercolor={.8 .8 1}, urlbordercolor={.8 .8 0}}
\usepackage[all]{hypcap}
\usepackage[export]{adjustbox}

\newcommand\ba{\begin{array}}
\newcommand\ea{\end{array}}
\newcommand\nn{\nonumber}

\newcommand\ri{\right}
\renewcommand\le{\left}

\newcommand{\feyn}[1]{#1\kern-0.45em/}

\renewcommand\a{\alpha}

\renewcommand\b{\beta}

\renewcommand\d{\delta}

\newcommand\D{\Delta}

\newcommand\m{\mu}

\renewcommand\th{\theta}

\newcommand\la{\langle}
\newcommand\ra{\rangle}

\newcommand\mc{\mathcal}

\newcommand{\mycolor}{black}

\begin{document}
\title{\textcolor{\mycolor}{The transition from homeostasis to stochasticity induced catastrophe}}
\author{Rudro R. Biswas}
\email{rrbiswas{@}purdue.edu}
\affiliation{Department of Physics and Astronomy, Purdue University, West Lafayette, IN 47907, USA}
\author{Charles S. Wright}
\affiliation{Department of Physics and Astronomy, Purdue University, West Lafayette, IN 47907, USA}
\affiliation{Monash Biomedicine Discovery Institute, Monash University, Melbourne, Australia}
\author{Kunaal Joshi}
\affiliation{Department of Physics and Astronomy, Purdue University, West Lafayette, IN 47907, USA}
\author{Srividya Iyer-Biswas}
\email{iyerbiswas{@}purdue.edu}
\affiliation{Department of Physics and Astronomy, Purdue University, West Lafayette, IN 47907, USA}

\begin{abstract}
What are the signatures of the onset of catastrophe? Here we present the rich system physics characterizing the transition from homeostasis to \textcolor{\mycolor}{stochasticity driven} breakdown in an experimentally motivated minimal model. Recent high-precision experiments on individual bacterial cells, growing and dividing repeatedly in a variety of environments, have revealed a previously unknown intergenerational scaling law which not only uniquely determines the stochastic map governing homeostasis, but also, as we show here, offers quantitative insights into the transition from the ``conspiracy principle'' regime (homeostasis) to the ``catastrophe principle'' regime and then to system breakdown. \textcolor{\mycolor}{In fact, upon closer examination, the stochastic map turns out to be a one-dimensional Kesten process;} these transitions occur as a single parameter\textcolor{\mycolor}{, the strength of the multiplicative noise term, is continuously tuned. Emergence of asymptotically scale invariant distributions with quantifiable power law tails, outlier driven extremal behavior and reverse monotonicity of the conditional exceedance distribution characterize this transition to catastrophe. In turn, prevention of rapid increase in the extremal event-driven rate of failure, in the interest of system preservation, causes the catastrophe regime to be strategically unfavorable.}
\end{abstract}
\maketitle

Catastrophes inspire awe and fear. While wildfires, collapsing bridges, epidemics, avalanches, oil spills, and stock market crashes may conjure dramatically different imagery, this apparent disparity masks underlying recurring themes: the randomness of the triggering event, the systemic breakdown, and the difficulty in restoring the complex system to its previously functional state~\cite{2016-miller}. These aspects of catastrophes present challenges to effectively modeling them and even to articulating precisely what constitutes a catastrophe. Grappling with the complexity of these systems (their interconnected parts, inherent probabilistic dynamics, and emergent behavior) while capturing the essential phenomenology in a tractable model presents part of the challenge. Another aspect lies in identifying the defining characteristic signatures that predict an impending catastrophe with sufficient precision to facilitate timely intervention before homeostasis is disrupted, rather than just mitigation after the fact.

\textcolor{\mycolor}{Homeostasis, the maintenance of constancy of some parameters of a complex (usually biological) system, is the antithesis of catastrophe.} Homeostasis is routinely conceptualized as a natural outcome of feedback control mechanisms~\cite{1961-wiener,2016-ramsay}. Within this perspective, the breakdown of homeostasis occurs due to feedback loops malfunctioning or becoming overwhelmed, leading to pathological states~\cite{2015-kotas,1951-wiener}. \textcolor{\mycolor}{In biological systems, different architectures can facilitate regulation of homeostasis, including integral feedback for perfect adaptation~\cite{2000-yi} and nonlinear feedback for dynamic compensation~\cite{2016-karin}. Disruption of homeostasis in specific physiological circuits has been proposed to underlie the development of diverse disease states~\cite{2000-topp,2002-el-samad,2018-frere,2020-korem-kohanim}.} Homeostatic systems can only recover from deviations confined to regions of phase space near the intended equilibrium or basin of attraction~\cite{2018-McGrath}. Such temporary disruptions occur when deviations remain within a recoverable range, and feedback mechanisms are delayed or partially effective, but eventually restore homeostasis (e.g., a mild allergic reaction where histamine release causes local swelling that is soon counteracted by anti-inflammatory signals). When deviations exceed a critical threshold, feedback mechanisms are either overwhelmed (e.g., in anaphylactic shock) or become positive feedback loops, further destabilizing the system, which can no longer self-correct without external intervention~\cite{2014-nijhout}. Whereas systems experiencing temporary disruptions typically exhibit a relatively fast rate of recovery once the perturbation is addressed (e.g., mild dehydration can be corrected by fluid intake), during a breakdown, the system either recovers very slowly or not at all without intervention (e.g., renal failure where dialysis is needed).



\begin{figure*}[t]
\begin{center}
\resizebox{0.95\textwidth}{!}{\includegraphics{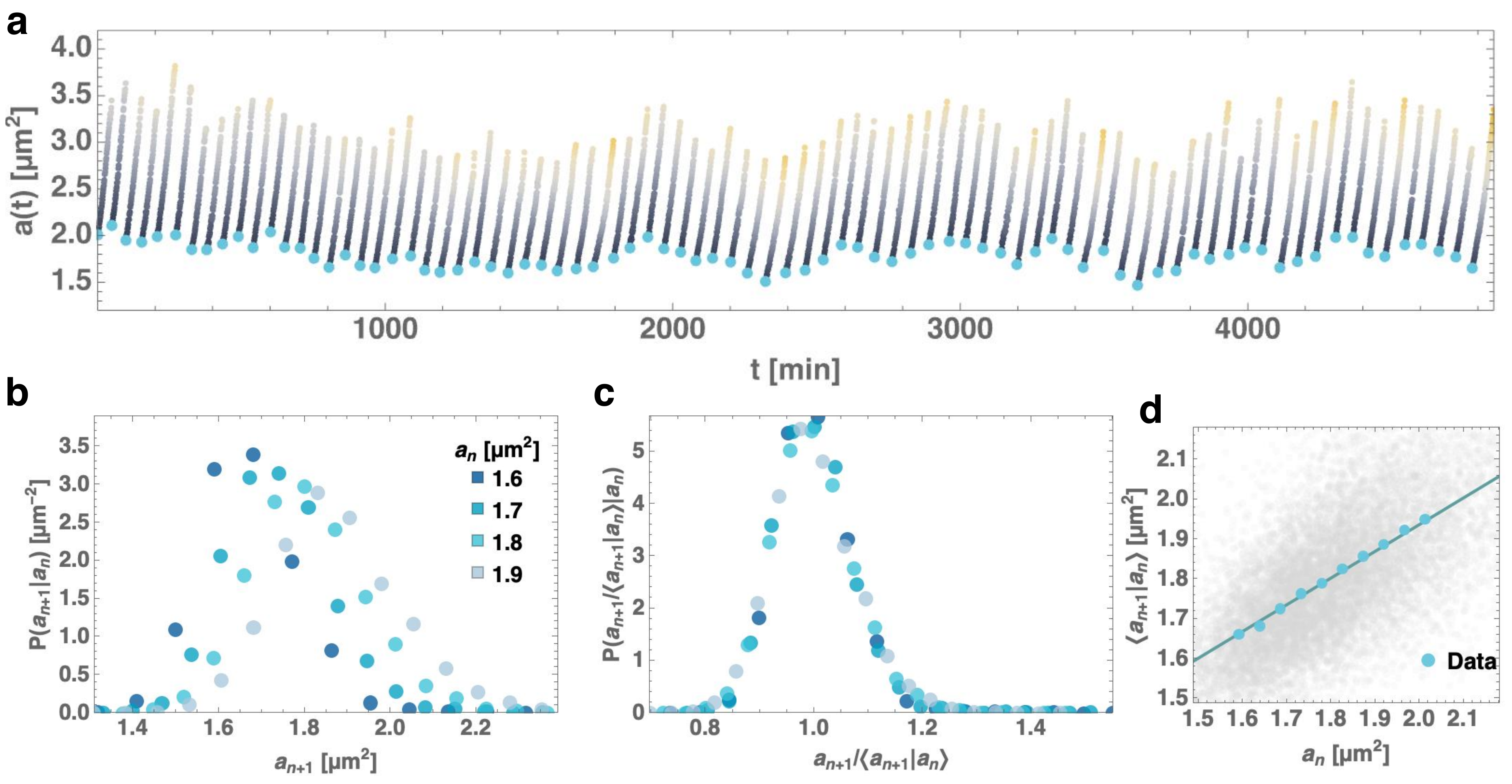}}
\caption{\textcolor{\mycolor}{\textbf{From experiments on living cells to a simple stochastic map equivalent to the Kesten process.} (a) shows experimentally obtained intergenerational cell size trajectories for a single cell. Each continuous curve shows a single cell generation, beginning at `birth' (i.e., just following the previous cell division) and terminating in a cell division process, following which the daughter cell is removed. From such trajectories, a characteristic cell size in every generation is selected: the initial sizes at `birth' as labeled by the blue dots. These depict a realization of the generational sequence of sizes ${a_{n}}$ used in the main text, where the subscript is the generation number. Such sets from a large number of identically growing and dividing cells are used for the statistical analysis in (b--d). (b) shows the different next-generation size ($a_{n+1}$) distributions resulting from a given initial size value of the current generation ($a_{n}$). The conditional means, $\la a_{n+1} | a_{n}\ra$, of these distributions are found to vary linearly with the given current generation initial size values, as shown in (d) where each blue dot represents the mean of a vertical sliver of the $(a_{n},a_{n+1})$ scatter plot, corresponding to a specific current generation initial size. When the distributions in (b) are mean-rescaled by the values in (d), a scaling collapse is found to occur, shown in (c). This distribution, independent of the current generation initial size, with unit mean (by definition) and a finite support, is the distribution $\Pi$ introduced in Eq.~\eqref{eq-scalinglaw}. Combining (c) and (d), we obtain the stochastic map Eq.~\eqref{eq-stochasticmap}, the starting point of our analysis, which upon appropriate variable transformation, is equivalent to the Kesten process, Eq.~\eqref{eq-kestenmap}. Data source: \cite{2024-architecture,2014-PNAS}.}}
\label{fig-introduction}
\end{center}
\end{figure*}

Recent high-precision longterm experiments~\cite{2023-emergentsimplicity,2024-mboc} on individual bacterial cells growing and dividing repeatedly in precisely controlled steady-state conditions have shown that (i) integenerational size evolution is Markovian, and (ii) revealed  a previously unknown intergenerational scaling law, which in turn specifies a new conceptual framework for stochastic homeostasis that is found to hold across  bacterial species, growth conditions and experimental modalities~\cite{2024-AnnuRevEmSim,2023-HomeostasisTheory,2024-architecture}. \textcolor{\mycolor}{(See Fig.~\ref{fig-introduction}.)} Briefly, denoting by $a_{n}$ the random variable newborn cell size at the beginning of the $n^{\text{th}}$ generation, this scaling law states that the mean-rescaled probability distribution of $a_{n+1}$, conditioned on the value of $a_{n}$, is independent of $a_{n}$ (for all $n$)  \textcolor{\mycolor}{(e.g., the scaling collapse in Fig.~\ref{fig-introduction}c, obtained by mean-rescaling the distinct distributions in Fig.~\ref{fig-introduction}b)}. Denoting this invariant distribution by $\Pi$ (with unit mean, by definition), and the conditional mean $\le\la a_{n+1}|a_{n}\ri\ra$ (i.e., the mean $\le\la a_{n+1}\ri\ra$, given a value of $a_{n}$) by the function $\m(a_{n})$, the scaling law can be represented thus:
\begin{align}\label{eq-scalinglaw}
\frac{1}{|\D s|}\text{Prob}\le(s < \frac{a_{n+1}}{\m(a_{n})}<s+\D s\ri) \stackrel{\D s \to 0}{=} \Pi(s).
\end{align}
Furthermore, over the physiological range accessible by experiments, $\m(a)$ is found to be well-approximated by a straight line~\cite{Jun2015}  \textcolor{\mycolor}{(Fig.~\ref{fig-introduction}d)}:
\begin{align}\label{eq-conditionalmean}
\m(a) = \a a + \b.
\end{align}
The function $\Pi(s)$ and constants $\a, \b$ can be obtained from high precision experimental data~\cite{2023-emergentsimplicity,2023-HomeostasisTheory,2024-mboc} \textcolor{\mycolor}{(Figs.~\ref{fig-introduction}c, d)}.

These experimental results above are precisely equivalent to the following formulation~\cite{2023-HomeostasisTheory,2024-architecture,2024-AnnuRevEmSim}: the intergenerational cell size trajectory evolves according to the following stochastic equation:
\begin{align}\label{eq-stochasticmap}
a_{n+1} = s_{n}(\a a_{n} + \b).
\end{align}
The $\le\{s_{n}\ri\}$ are independent random variables drawn from the same distribution $\Pi$ (with unit mean) defined in Eq.~\eqref{eq-scalinglaw} and represent the source of (multiplicative) noise. In what follows, we will denote the probability distribution of $a_{n}$ by the functions $P_{n}$. Eq.~\eqref{eq-stochasticmap} is difficult to solve for an arbitrary functional form of $\Pi(s)$ except for the case $\a=0$, when the size distribution in any generation $n$ after the initial one is exactly
\begin{align}\label{eq-perfecthomeostasis}
P_{n}(a) = \Pi(a/\b)/\b \;\; (\a = 0).
\end{align}
Thus, when $\a=0$, the size distribution function reaches an initial size-independent steady state in one generation (perfect homeostasis) with mean-rescaled distribution $\Pi$.

The stability of size distributions arising from the random walk given by Eq.~\eqref{eq-stochasticmap} was analyzed in~\cite{2023-HomeostasisTheory}. Of particular interest in~\cite{2023-HomeostasisTheory} was the existence of (cell size) homeostasis, which is the initial condition-independence and well-behaved nature of the size distribution after many cell generations (i.e., the longterm size distribution).  As we recapitulate  below, this analysis revealed a sequence of conditions, Eq.~\eqref{eq-stabilityconds}, involving $\a$ (from Eq.~\eqref{eq-stochasticmap}) and moments $m_{k}$ of the probability distribution $\Pi$ (Eq.~\eqref{eq-momentfunc}), which simultaneously govern the longterm convergence and initial condition-independence of the moments of the longterm size distribution. The meeting of all these conditions  is equivalent to a single geometric relation, Eq.~\eqref{eq-homeostasiscond}, between $\a$ and the upper limit, $s_{\text{max}}$, of the extent of the distribution $\Pi$. Under this condition the longterm size distribution is homeostatic, i.e., well-behaved and initial condition-independent with convergent moments. 

\begin{figure*}[t]
\begin{center}
\resizebox{0.95\textwidth}{!}{\includegraphics{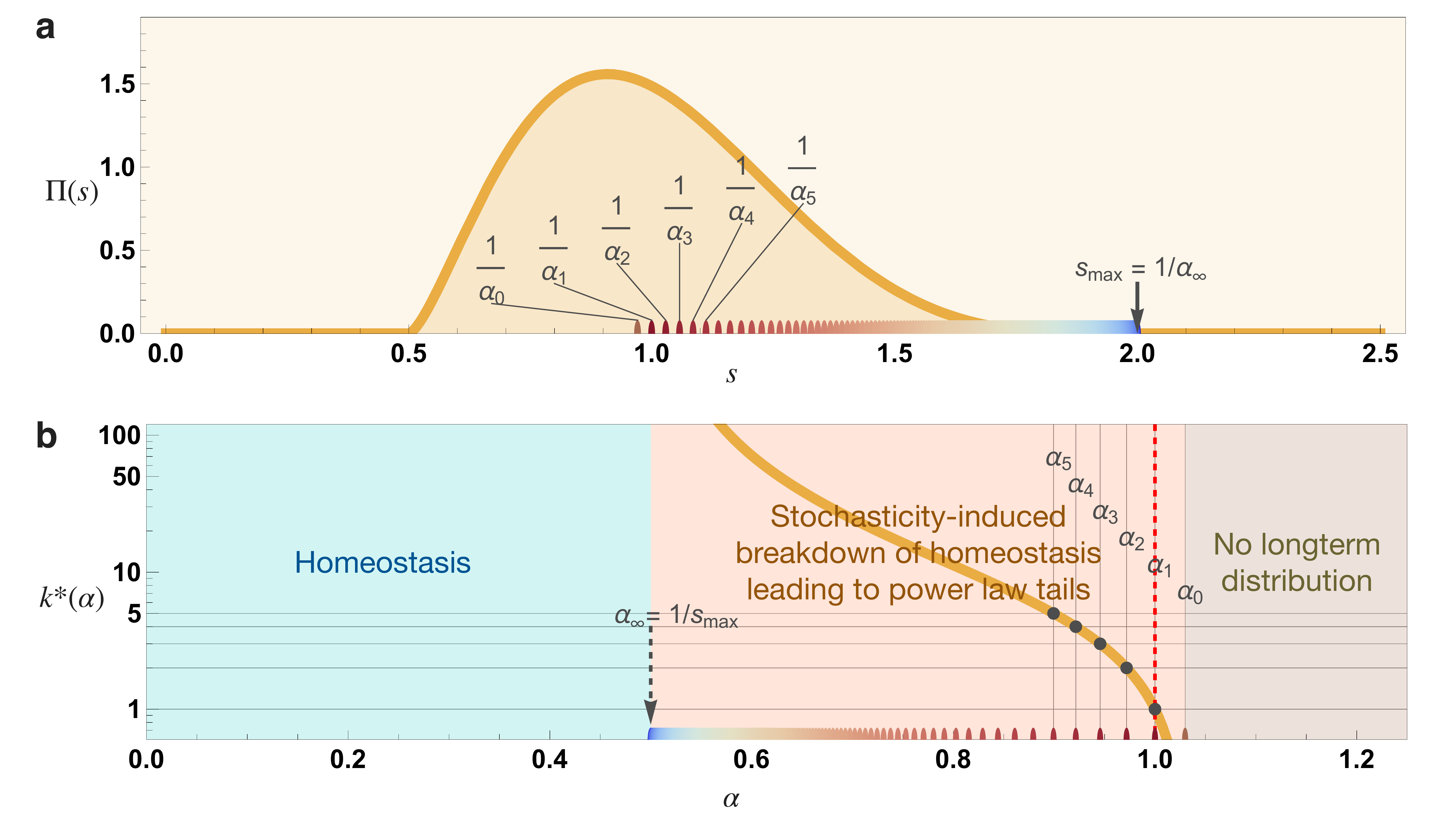}}
\caption{\textbf{Stochasticity governs specification of distinct regimes corresponding to: homeostasis, onset of catastrophe, and system breakdown.} (a) Shows the model $\Pi(s)$ used for simulations of Eq.~\eqref{eq-stochasticmap}. It is a Beta distribution with parameters $(2.5,5)$ in standard notation, stretched and translated to lie between $s_{\text{min}}=0.5$ and $s_{\text{max}}=2$. The mean is $1$ as required. (b) Shows three different dynamical regimes that occur as $\a$ is varied in Eq.~\eqref{eq-stochasticmap}. $\a=0$ corresponds to perfect homeostasis of the state variable $a$. The region of stable homeostasis extends between $0\leq \a < \a_{\infty}  = 1/s_{\text{max}} = 0.5$, where all moments of the long time $a$-distribution are finite, independent of initial conditions and the distribution itself is characterized by strict bounds, Eq.~\eqref{eq-amaxmin}. The regime of stochasticity-induced breakdown of homeostasis extends over $\a_{\infty} < \a < \a_{0}$, with $\a_{0}>1$ (Eq.~\eqref{eq-alpha0}), where a longterm $a$-distribution exists but is heavy-tailed. The distribution here is characterized by a scale-invariant power law tail decaying with an $\a$-dependent exponent $-(k^{*}(\a)+1)$ (see Figs.~\ref{fig-tails} and \ref{fig-trajectories}). All moments of order $k^{*}(\a)$ and higher are undefined. $k^{*}(\a)$ is obtained by solving $\a_{k}=\a$ for $k$ (Eq.~\eqref{eq-paretoexponent}) and is plotted as the orange curve in (b) -- this solution exists only within the stochasticity-induced breakdown regime $\a_{\infty} < \a < \a_{0}$, where it decreases monotonically from $+\infty$ to $0$. The values $\a_{k}$ (Eq.~\eqref{eq-definealphap}) corresponding to integer values of $k$ are indicated on the $\a$-axis in (b), and the corresponding values of $1/\a_{k}$ are indicated in (a). By comparing (a) and (b), it is clear that when $1/\a$ lies within the tail region of $\Pi(s)$ in (a), the value of $k^{*}(\a)$ increases rapidly and the size distribution power law tail becomes almost undetectable. When $\a$ crosses $\a_{1}=1$ (dashed red line), even the mean becomes undefined and this is also when dynamical instability sets in from the deterministic viewpoint. However, the longterm distribution persists till $\a$ reaches another point $\a_{0}\simeq 1.03$, beyond which lies the third and final dynamical regime with no defined longterm distribution. \textcolor{\mycolor}{A second example of $\Pi(s)$, and corresponding characterizations of its three regimes, is presented in Fig.~\ref{fig-alphafigB}.}}
\label{fig-alphafig}
\end{center}
\end{figure*}

Our approach above \textcolor{\mycolor}{culminating in Eq.~\eqref{eq-stochasticmap}} captures the homeostasis regime's phenomenology precisely---as observed in a naturally occurring complex and adaptive system---and does so in a minimal, analytically tractable model with a \textcolor{\mycolor}{ simple} tunable parameter \textcolor{\mycolor}{$\a$}. \textcolor{\mycolor}{ In this work we} examine what this minimal model reveals about the quantitative signatures of the onset of catastrophe as \textcolor{\mycolor}{this} parameter is tuned. While even simple bacterial cells have many additional layers of control to prevent routine catastrophes, examining the model in the regime where homeostasis is no longer guaranteed provides rich insights into the system physics governing the stochasticity induced transition to catastrophe. Fig.~\ref{fig-alphafig}(b) captures and provides a preview of the rich physics of homeostasis breakdown revealed in this work. The richness of the phenomenology contrasts with the widely-held assumption that cell size homeostasis merely requires that the condition $\a<1$ be met, as deduced previously from straightforward theoretical exercises that have considered many heuristic models, both  deterministic and random walks, governing intergenerational cell size evolution~\cite{Willis2017,Modi2017,lin2017effects,logsdon2017parallel,sauls2016adder,2015-JunReview,deforet2015cell,Jun2015,amir2014cell,Spiesser2012,2000-nurse,1987-tyson,1981-nurse}.

The question we focus on in this work can be reformulated as follows: what phenomenologies arise when the strength of the multiplicative noise is increased beyond the threshold beyond which homeostatis breaks (i.e., when Eq.~\eqref{eq-homeostasiscond} is violated)? \textcolor{\mycolor}{In the absence of exact solutions to Eq.~\eqref{eq-stochasticmap} for generic $\Pi(s)$,} progress is made possible by making a connection with extant literature on the Kesten process~\cite{1973-kesten}, which is related to the one in Eq.~\eqref{eq-stochasticmap} through a variable transformation.  The Kesten process, since it was first introduced by Kesten~\cite{1973-kesten} decades ago, has re-appeared in diverse contexts involving complex systems, including earthquakes~\cite{2023-earthquake}, stochastic optimization algorithms~\cite{1999-algorithms}, economic and financial markets~\cite{2003-finance-book,2023-finance}, socioeconomic behaviors~\cite{1999-zipf,2024-socioeconomic}, and social media usage~\cite{2023-socialmedia}. \textcolor{\mycolor}{It has recently appeared in traditional statistical physics problems in the context of stochastic resetting~\cite{2023-gueneau,2024-biroli}. A detailed survey of the Kesten process and associated literature, including active research topics, is available as~\cite{2016-buraczewski}.}

Kesten's process~\cite{1973-kesten} is a stochastic map of the form:
\begin{align}\label{eq-kestenmap}
Y_{n+1} = M_{n} Y_{n} + Q,
\end{align}
where $M_{n}$ form a series of i.i.d. random variables and $Q>0$. \textcolor{\mycolor}{(We will assume intergenerational constancy of $Q$ for brevity even though many general results below continue to hold when $Q$ is a light-tailed positive random variable.)} Using fundamental results from the theory of multiplicative noise, Kesten proved \cite{1973-kesten} that the longterm distribution exists only when the statistics of the $M_{n}$ variables ensure that
\begin{align}\label{eq-kestencond}
\le\la \ln M \ri\ra < 0.
\end{align}
Since Eq.~\eqref{eq-stochasticmap} can be recast as Eq.~\eqref{eq-kestenmap} with the identifications $Y_{n} = a_{n} + \b/\a$, $M_{n} = \a s_{n}$, and $Q = \b/\a$, we conclude that the condition for existence of a longterm probability size distribution, Eq.~\eqref{eq-kestencond}, becomes for our process, Eq.~\eqref{eq-stochasticmap},
\begin{align}\label{eq-stabilityconds0}
\le\la \ln \a s \ri\ra < 0, \text{ i.e., } \a < e^{-\la \ln s \ra}.
\end{align}
\textcolor{\mycolor}{Furthermore, Kesten deduced that the process in Eq.~\eqref{eq-kestenmap} generically leads to steady state distributions with power law tails when the condition Eq.~\eqref{eq-kestencond} is satisfied. Subsequently, a method was discovered to calculate the exponent of this power law~\cite{1991-goldie}: if the distribution of $M$ in Eq.~\eqref{eq-kestenmap} is such that the following equation in $k$:
\begin{align}
\le\la M^{k} \ri\ra = 1,
\end{align}
is satisfied by $k=k^{*}$, then the power law exponent of the tail of the steady state probability distribution (the density function) is $(k^{*}+1)$. The presence of this power law tail also ensures that moments of the steady state distribution of $Y$, $\lim_{n\to\infty}\le\la Y_{n}^{p}\ri\ra$ for positive exponents $p$, are finite only when the exponent $p < k^{*}$. What is remarkable about these results is that such power law tails arise, i.e., $Y$ has a `heavy-tailed' distribution, even though the distribution for $M$ may be light-tailed.}

\textcolor{\mycolor}{In what follows, we re-derive quantitatively how the power law tails emerge in Eq.~\eqref{eq-stochasticmap}, paying special attention to a special feature of our problem, i.e., that $\Pi(s)$ has finite support in the range $(s_{\text{smin}}, s_{\text{max}})$. This feature leads to clearly defined regime of homeostasis, as has been observed in the absence of stress in living bacterial cells (Fig.~\ref{fig-introduction}), and regimes of stochasticity-induced homeostasis breakdown (the `catastrophe' regime) and complete breakdown. We then analyze how `catastrophic' consequences arise as the multiplicative noise strength is increased by increasing $\a$ beyond the homeostasis-breaking threshold, using specific experimentally-motivated examples and sophisticated tools, including those from extant literature on actuarial sciences and heavy-tailed distributions~\cite{resnick2007heavy,2003-finance-book,MANDELBROT20031,mandelbrot1982fractal,2022-catastrophe}. Finally, increasing the noise strength beyond the ultimate limit identified by Kesten in Eq.~\eqref{eq-kestencond} (which we show occurs at a value for which $\a>1$), a well-defined longterm distribution can no longer be accessed, leading to complete breakdown. The random walk represented by the stochastic process of Eq.~\eqref{eq-stochasticmap} thus offers an experimentally motivated theoretical model for investigating the stability of stochastic processes as $\a$ is tuned one-dimensionally from homeostasis to catastrophe to complete breakdown.}

\textcolor{\mycolor}{What we present here, motivated by exciting recent experimental developments, is a unique synthesis of concepts, frameworks, terminology and results from seemingly disparate fields, spanning biology, stochastic processes, complex systems, actuarial sciences, statistical risk management, quantitative finance and control engineering. We hope to offer enough new motivation, concepts and results for the aficionados, while also providing the generalist with a template and unified language for approaching similar problems whether they arise in other contexts in biology or in other complex systems.}

\emph{Notation:} {\textcolor{\mycolor}{From here on} we  \textcolor{\mycolor}{will} refer to \textcolor{\mycolor}{the} stochastic variables $a_{n}$ as the `system state'  or `state' variables, identifying them with cell size only in contexts relevant to the biological motivation described above. \textcolor{\mycolor}{Our observations and deductions can be straightforwardly translated to the myriad scenarios where homeostasis occurs and Kesten processes are found to arise.}}

\subsection*{Results}
\subsubsection*{\textbf{Part I: Emergence of distinct regimes of stochastic dynamics.}}
\textcolor{\mycolor}{We now analyze the steady state distribution arising from Eq.~\eqref{eq-stochasticmap} following Kesten, Goldie and subsequent analyses ~\cite{1973-kesten,1991-goldie,2016-buraczewski}, paying attention to a new feature of our problem, namely that $\Pi(s)$ has finite support in the range $(s_{\text{smin}}, s_{\text{max}})$. We show that three distinct stochasticity-induced dynamical phases arise depending on the value of $\a$ and the properties of $\Pi(s)$, the distribution of the stochastic variables $s_{n}$. Anticipating their usefulness later on, we  define two quantities related to $\Pi$.} The first is a general moment function of the distribution $\Pi$:
\begin{align}\label{eq-momentfunc}
m_{p} = \le\la s^{p} \ri\ra = \int_{0}^{\infty}  s^{p}\Pi(s) \, ds.
\end{align}
The argument, $p$, of the function is the index of the moment and we can take it to be continuously varying and satisfying $p\geq 0$. We will also make the physiologically reasonable assumption that $\Pi(s)$ is nonzero within a finite range $s_{\text{min}} < s < s_{\text{max}}$, with $s_{\text{min/max}}$ chosen such that the  range is the minimum possible. \textcolor{\mycolor}{(See Fig.~\ref{fig-introduction}c for an experimental example.)} Thus, $m_{0} = m_{1} = 1$ (normalized probability and unit mean) and as $p\to\infty$, $m_{p} \to (s_{\text{max}})^{p}$ from below, i.e., $m_{p} < (s_{\text{max}})^{p}$ always. The related second definition is of the function:
\begin{align}\label{eq-definealphap}
\a_{p} &= \le(m_{p}\ri)^{-1/p}, \; p>0,\nn\\
&= \lim_{p\to 0+} \le(m_{p}\ri)^{-1/p}, \; p=0.
\end{align}
$\a_{p}$ is a strictly decreasing function of $p$, as can be seen from the following argument. Consider two positive numbers, $p_{1}>p_{2}$, and assume that $s$  has nonzero variance (otherwise $s^{p} = 1$ for all $p$ since the mean is one). Since $x^{p_{1}/p_{2}}$ is a convex function of $x$ for $x>0$, using Jensen's inequality, $\le\la s^{p_{1}}\ri\ra = \le\la \le(s^{p_{2}}\ri)^{p_{1}/p_{2}}\ri\ra >  \le(\le\la s^{p_{2}}\ri\ra\ri)^{p_{1}/p_{2}}$ and so $\a_{_{p_{1}}} < \a_{p_{2}}$.

It is straightforward to see that 
\begin{subequations}
\begin{align}
\a_{1} &= 1, \; \text{($\Pi$ has unit mean)}\\
\a_{\infty} &= \lim_{p\to \infty} \a_{p} = 1/s_{\text{max}}, \; \text{and}\\
\a_{0} &= \lim_{p\to 0+} \le(m_{p}\ri)^{-1/p} = e^{- \la \ln(s)\ra}>\a_{1} =1. \label{eq-alpha0}
\end{align}
\label{eq-alphalimits}
\end{subequations}

We can now pithily describe the stability conditions governing Eq.~\eqref{eq-stochasticmap}. \textcolor{\mycolor}{As can be deduced from explicit evaluation of the intergenerational evolution of integer moments of Eq.~\eqref{eq-kestenmap} (as in \cite{2009-alsmeyer} when analyzing perpetuities following the Kesten process) or of Eq.~\eqref{eq-stochasticmap} (as in \cite{2023-HomeostasisTheory} the context of intergenerational cell size dynamics), all integer moments of the state variable, $\la (a_{n})^{k'}\ra$, with exponents $k'\leq k$ reach finite and initial condition-independent longterm values, $\la a^{k'} \ra_{\infty}$, if and only if}
\begin{align}\label{eq-stabilityconds}
\a < \a_{k}.
\end{align}
Furthermore, we will describe later below that this condition along with its implications for convergence of moments hold for all real $k$ over the continuous range $0\leq k < \infty$ and not only when $k$ is a positive integer.

As $k\to\infty$, we obtain the necessary and sufficient condition for all state variable moments to converge to finite initial condition-independent values and ensure existence of a well-behaved initial condition-independent homeostatic distribution. \textcolor{\mycolor}{This was shown to be $P(|A|<1)=1$ in the context of the Kesten process~\cite{2009-alsmeyer,2016-buraczewski}. The equivalent result in the context of Eq.~\eqref{eq-stochasticmap} is the following \cite{2023-HomeostasisTheory}:}
\begin{align}\label{eq-homeostasiscond}
\a \leq \a_{\infty}  = 1/s_{\text{max}}.
\end{align}
(The equality sign arises from the fact that the $\a_{p}$ are always less than their $p\to\infty$ limit, $\a_{\infty}$.) This condition is remarkably consistent with experimental observations of cell size distributions of bacterial cells in homeostasis across growth conditions, experimental modalities and species~\cite{2023-HomeostasisTheory,2023-emergentsimplicity,2024-mboc,2024-AnnuRevEmSim}.

\textcolor{\mycolor}{Remarkably, the stability condition Eq.~\eqref{eq-stabilityconds} is also applicable to the stable existence of the $k=0$ moment in the long term, which corresponds to the very existence of a normalizable longterm probability distribution. This can be deduced from combining Kesten's seminal result~\cite{1973-kesten}, Eqs.~\eqref{eq-kestencond} and ~\eqref{eq-stabilityconds0}, with the definition in Eq.~\eqref{eq-alpha0}, which yields the stability condition for exponential convergence to a steady state distribution to be:}
\begin{align}\label{eq-stabilitycond0}
\a < \a_{0}.
\end{align}

These stability conditions are visually presented as color-coded regimes in Fig.~\ref{fig-alphafig}(b). When $0\leq \a < \a_{\infty} = 1/s_{\text{max}}$ (blue), there exists a longterm distribution that is stable, initial condition-independent and with finite moments. Increasing $\a$ further, $\a$ successively crosses the values $\a_{k}$ with decreasing values of $k$, such that when it lies in the range $\a_{k} \leq \a < \a_{k-1}$, the $k^{\text{th}}$ and higher order moments of $a$ are indeterminate, while all lower order moments reach finite and initial condition-independent longterm values. In the entire range $\a_{\infty}<\a<\a_{0}$, a longterm distribution is still reached and, as shown next, is characterized by the emergence of a scale-invariant power law tail. We call this the stochasticity-induced breakdown regime (red). Finally, when $\a$ increases beyond $\a_{0}$ (which is greater than $1$), the concept of a normalizable longterm distribution needs to be abandoned entirely due to runaway dynamics when all state variable values go to infinity (brown).

\subsubsection*{\textbf{Part II: Catalog of signatures that distinguish systems in stable homeostasis and those which are in the catastrophe-prone regime}}

{\bf \emph{Emergent scale invariance in the stochasticity-induced breakdown regime.} }What universal signature characterizes the region $\a_{\infty}<\a<\a_{0}$ (the `stochasticity-induced breakdown' phase in Fig.~\ref{fig-alphafig}(b)) where a longterm distribution exists yet whose moments higher than a certain order are all undefined/infinite? Consider anew the observation from Eq.~\eqref{eq-stabilityconds} that when $\a_{k}\leq\a<\a_{k-1}$, all $k^{\text{th}}$ and higher moments of $a$ are undefined~\cite{2009-alsmeyer,2023-HomeostasisTheory}. A possible mechanism for this to occur is that the $a \to \infty$ asymptotic behavior of the long time distribution has a power law tail which falls off slower than $a^{-(k+2)}$ but faster than $a^{-(k+1)}$. This emergent scale-invariance of the tail of the state variable distribution is the defining characteristic of the stochasticity-induced breakdown phase. This scale invariance is an emergent feature since the input stochastic distribution $\Pi(s)$, that of $s_{n}$ in Eq.~\eqref{eq-stochasticmap}, has a finite range and no power law or any other type of heavy tail. Indeed, it has been shown that when the condition in Eq.~\eqref{eq-kestencond} is met, the long time asymptotic distribution of $Y$ (in Eq.~\eqref{eq-kestenmap}, related to $a$ by a linear transformation) has a power law tail~\cite{1973-kesten}. Furthermore, the exponent of this power law tail can also be calculated~\cite{1991-goldie} as a function of $\a$, as we \textcolor{\mycolor}{recalculate} below.  (The possibility that the state variable distribution may have no \textcolor{\mycolor}{power law} tail at all is mathematically allowed \textcolor{\mycolor}{yet seldom considered} \textcolor{\mycolor}{(\cite{2009-alsmeyer} is a notable exception)}; yet it is this homeostasis regime that \textcolor{\mycolor}{is of most interest in the cell} size regulation problem~\cite{2024-AnnuRevEmSim}.)

 Assume the form $P_{\infty}(a) \simeq N_{\infty}/a^{(1+k^{*}(\a))}$ for the tail of the longterm distribution of $a$, characterized by an $\a$-dependent exponent and overall coefficient $N_{\infty}$.  Far into this tail ($a\to\infty$), the presence of $\b$ can be ignored in Eq.~\eqref{eq-stochasticmap}, which can be replaced by $a_{n+1}\simeq \a s_{n} a_{n}$. At sufficiently large values, both $a_{n+1}$ and $a_{n}$ fall within the power law tail region of the probability distribution and so:
\begin{align}
P_{\infty}(a_{n+1}) &\simeq \int ds \Pi(s) \int da_{n} P_{\infty}(a_{n}) \d(a_{n+1} - \a s a_{n})\nn\\
&= \int ds \frac{\Pi(s)}{\a s} P_{\infty}\le(\frac{a_{n+1}}{\a s}\ri).
\end{align}
Substituting the power law behavior of the tail, $P_{\infty}(a) \simeq N_{\infty}/a^{(1+k^{*}(\a))}$, we find that $k^{*}(\a)$ satisfies:
\begin{align}\label{eq-paretoexponent}
\a^{k^{*}(\a)} \int ds \Pi(s) s^{k^{*}(\a)} = 1.
\end{align}
In other words, $k^{*}(\a)$ is simply the solution (for $k$) to the equation $\a_{k} = \a$, where $\a_{k}$ is as defined in Eq.~\eqref{eq-definealphap}! Consider now the straightforward fact that the power law tail $\simeq N_{\infty}/a^{(1+k^{*}(\a))}$ leads to undefined moments $\le\la a^{p}\ri\ra$ for all real $p \geq k^{*}(\a))$. This fact and the physics underlying Eq.~\eqref{eq-stabilityconds0}~\cite{1973-kesten} together show that the relationship between the inequalities in Eq.~\eqref{eq-stabilityconds} and finiteness of certain moments of $a$, derived \textcolor{\mycolor}{above} only for positive integer values of $k$, are valid for all real $k$ within the continuous range $0\leq k \leq \infty$. \textcolor{\mycolor}{For more details, nuances and history of these derivations, we direct the reader to \cite{2016-buraczewski}.}

\begin{figure*}[t]
\begin{center}
\resizebox{0.99\textwidth}{!}{\includegraphics{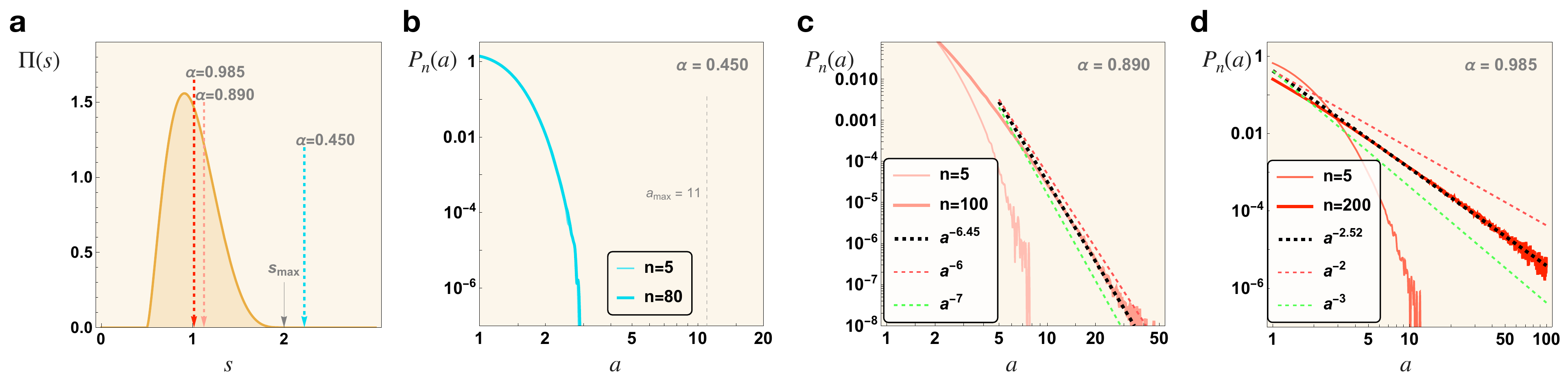}}
\caption{\textbf{Emergence of heavy-tailed Pareto power law distributions in the stochasticity induced onset of catastrophe regime.} (a) The stretched Beta distribution form used for $\Pi(s)$ in our simulations (Fig.~\ref{fig-alphafig}(a) has more details). The locations of $1/\a$ for the three $\a$ values used in (b-2) are indicated by arrows. (b-d) Numerically generated evolution of state variable distributions shown on log-log scale, from generation $n=5$ to $n=80$ (b), $100$ (c) or $200$ (d) (the longterm limit; \textcolor{\mycolor}{see Fig.~\ref{fig-convergence} for evaluation of convergence to steady state}), starting from initial condition $a_{0}= 0.8$ (b) or $1.0$ (c,d). The `target'/longterm average of $a$ has been chosen to be unity in all plots. 10-100 million trajectories were simulated for each figure. (b) $\a = 0.45 < 0.5 = 1/s_{\text{max}}$ and so strict homeostasis occurs with finite, well-defined initial-condition-independent moments of the long time state variable distribution. There is an upper cutoff, $a_{\text{max}}$, that can be analytically calculated (Eq.~\eqref{eq-amaxmin}) and is consistent with the numerical result. (c) $\a$ is now increased to satisfy $\a_{6}<\a<\a_{5}$. Thus $k = 6$ (see main text) and the $a$-distribution is seen to develop a power law tail with exponent $-(k^{*}(0.89)+1) \simeq -6.45$ lying between $-7$ and $-6$. In this case, the sixth and higher moments of the longterm distribution are undefined and initial-condition-dependent. (d) $\a$ is now large enough that $\a_{2}<\a <1$ and so even the second moment of the longterm $a$-distribution does not exist. As expected, the heavy tail has a power law decay with the exponent $-(k^{*}(0.985)+1) \simeq -2.52$ lying between $-3$ and $-2$.}
\label{fig-tails}
\end{center}
\end{figure*}

Due to the previously \textcolor{\mycolor}{argued} monotonic decreasing property of $\a_{p}$ as a function of $p$, if $\a_{k+1}\leq\a<\a_{k}$, then we also have $k<k^{*}(\a)\leq k+1$. As $\a$ is increased starting from $0$ (perfect homeostasis) but to a value less than $\a_{\infty} = 1/s_{\text{max}}$, the system continues to be in homeostasis and Eq.~\eqref{eq-paretoexponent} does not have a solution. Homeostasis is broken as $\a$ crosses $\a_{\infty}$ and   Eq.~\eqref{eq-paretoexponent} becomes solvable (Fig.~\ref{fig-alphafig}(b)) with $\lim_{\a\to\a_{\infty}+}k^{*}(\a) = +\infty$. As $\a$ increases towards $\a_{1}=1$, the point at which the system is no longer stable according to the deterministic arguments,  $k^{*}(\a)$ drops from $+\infty$ to $1$, the power law tail becomes heavier and more and more of the lower moments of the size distribution (with orders $\geq k^{*}(\a)$) become undefined. At $\a=1$, the longterm distribution has a tail that falls off as $a^{-2}$, precluding the existence of a finite and initial condition-independent mean. Yet as $\a$ increases further, the longterm distribution continues to exist, though devoid of finite moments at any positive integer order. This remnant of describability using the concepts of stochastic processes also disappears when $\a$ approaches $\a_{0}$ (Eq.~\eqref{eq-alpha0}) from below and the distribution tail approaches $a^{-1}$, which is not normalizable. For $\a>\a_{0}$ there is no longterm limit for the state variable distribution to converge to as all values run away to infinity. This description of the evolution of distribution \emph{shape} with $\a$ mirrors the moment-based description provided earlier (also, Fig.~\ref{fig-alphafig}).

\begin{figure*}[t]
\begin{center}
\resizebox{14cm}{!}{\includegraphics{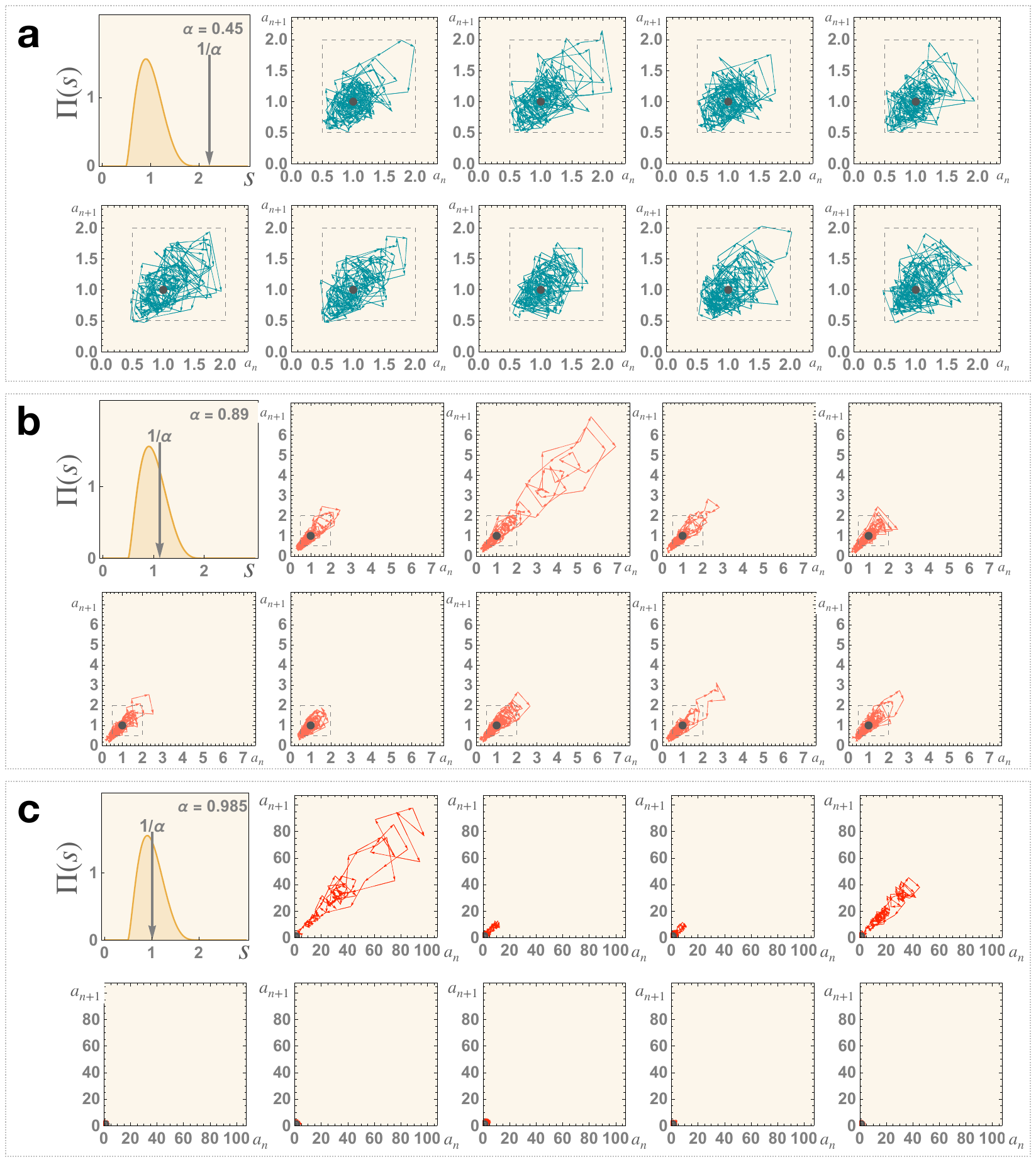}}
\caption{\textbf{In the stochasticity induced onset of catastrophe regime giant excursions away from the ``set point value'' are frequently observed as is typical of the heavy tailed process governing the dynamics in this regime.} In each of Fig.~(a-c), the first mini-plot provides a visualization for the $\a$-value used in the subfigure by placing $1/\a$ on the $\Pi(s)$ graph (same Beta function as used in Fig.~\ref{fig-alphafig}(a)).  The remaining nine mini-plots show independent state variable ($a$) trajectories as a succession of points $(a_{n},a_{n+1})$ with increasing $n$, the generation number. Thus, their axes show the next generation $a$ against the current generation value. These axes are normalized by the `target'/asymptotic average value, $\le\la a\ri\ra_{\infty}$ to facilitate comparison between Figs.~(a-c). This target value is also indicated by the gray dot. Each trajectory contains $200$ generations. The dashed rectangle is a guide to the eye showing the extent of excursions allowed by variation in $s$ from the target point -- as homeostasis is violated, stochastic excursions outside this range becomes more likey. It corresponds to the range $(s_{\text{max}}, s_{\text{min}})$ of $\Pi(s)$, the exact boundary limiting excursions when $\a=0$ (see Eq.~\eqref{eq-perfecthomeostasis}). (a) When the homeostasis condition, Eq.~\eqref{eq-homeostasiscond}, is satisfied (same as in Fig~\ref{fig-tails}(b)), the $a$ values circulate within/around the marked boundary and have the same order of magnitude as when perfect homeostasis occurs at $\a=0$. Large excursions are extremely improbable (suppressed exponentially or faster\textcolor{\mycolor}{, since the distribution of $a$ is `light-tailed'}). (b) and (c) show sample trajectories when $\a$ violates the stochastic homeostasis condition but not the deterministic stability condition, i.e., $\a<1$. The $\a$-values used in Figs.~(b) and (c) correspond to the those in Figs~\ref{fig-tails}(c) and (d), respectively. These now show intermittent giant excursions well outside the marked boundaries. These `black swan' events characterize heavy tail phenomena, their probabilities falling off slowly as a power law in extent. While there is no characteristic scale describing black swans in this `scale-invariant' regime, their observed sizes increase with $\a$ due to thickening and slowing of the power law tail of the longterm $a$ distribution, Fig.~\ref{fig-alphafig}(b), as occurs here when going from Fig.~(b) to (c) (note the axis ranges).}
\label{fig-trajectories}
\end{center}
\end{figure*}

The asymptotic tail of the survival function (one minus the cumulative function, the complementary cumulative distribution function) is thus also a power law, the so-called Pareto power law, with `Pareto exponent' equal to $-k^{*}(\a)$ (Eq.~\eqref{eq-paretoexponent}) in our model. Distributions with such power law tails are of widespread interest and are an important subset of the class of `heavy-tailed' distributions with remarkable non-intuitive properties which can pose a challenge to conventional strategies for mitigating risk~\cite{taleb2020statistical,2003-finance-book}. We have numerically verified the emergence of heavy, power law-tailed longterm distributions from Eq.~\eqref{eq-stochasticmap} whose Pareto exponents can be tuned by changing the value of $\a$ according to Eq.~\eqref{eq-paretoexponent}. For simulations we have used a stretched Beta distribution form for $\Pi(s)$ as shown in Fig~\ref{fig-alphafig}(a), whose range extends between $s_{\text{min}} = 0.5$ and $s_{\text{max}}=2$. \textcolor{\mycolor}{(For an alternate example, see Fig.~\ref{fig-alphafigB}. We chose a  representative parametrizable, analytic, smooth distribution which has finite extent, a bell shape, and allowed nonzero skewness: all generic features also common to the experimentally measured distributions in the cellular context, Fig.~\ref{fig-introduction}(c).)} To facilitate comparison between different $\a$, we have used mean-rescaled values for $a$, dividing them by the mean $\le\la a\ri\ra_{\infty} = \b/(\a-1)$. The decreasing series formed by $\a_{k}$ (with increasing $k$) starts at $1$ for $k=1$ and decreases to $\a_{\infty}=1/s_{\text{max}} = 0.5$ as $k\to \infty$. Specific values of $\a_{k}$ corresponding to the distribution in Fig~\ref{fig-alphafig}(a) are: $\a_{0} \simeq 1.03$, $\a_{1} = 1$ (by definition), $\a_{2} \simeq 0.972, \a_{3} \simeq 0.946, \ldots , \a_{5} \simeq 0.899, \a_{6} \simeq 0.879, \ldots$ (Fig~\ref{fig-alphafig}(b)). Thus, in Fig~\ref{fig-tails}(b), well-behaved homeostasis occurs for $\a = 0.45 <\a_{\infty} = 0.5$, when the mean-rescaled longterm distribution is also bounded by a maximum limit $a_{\text{max}}/\le\la a\ri\ra_{\infty} = (1-\a)/(1-\a s_{\text{max}}) = 11$ (see Eq.~\eqref{eq-amaxmin} below). In the stochasticity-induced breakdown regime, when $\a_{k+1}<\a<\a_{k}$ for integer $k$, we expect a power law tail in the longterm distribution with the exponent $-(k^{*}(\a)+1)$ lying in the range $-(k+1)$ and $-(k+2)$. Fig~\ref{fig-tails}(c) and (d) show the cases when $\a=0.89$ and $0.985$, for which $k^{*}(0.89) \simeq 5.45$ ($k=5$) and $k^{*}(0.985) \simeq 1.52$ ($k=1$) respectively. It is clear from the figures that the numerically generated power law tails in the longterm distributions (after $\sim 100$ generations) have exponents that match the derived values of $-(k^{*}+1)$. Furthermore, from our simulations it is clear that as $\a$ decreases towards $\a_{\infty} = 1/s_{\text{max}}$ and $k^{*}(\a)$ becomes large, these distribution tails fall off sharply, carry lesser total probability weights and become harder to detect. This occurs simultaneously with the value of $1/\a$ moving out across the tail of the $\Pi(s)$ distribution, as can be seen from comparing the subfigures in Fig~\ref{fig-alphafig}. Thus, within the bounds of experimental detection, a homeostatic distribution is guaranteed if $1/\a$ is placed outside to the right of the main bulk of the measured $\Pi$-distribution, in agreement with the stricter theoretical condition in \cite{2023-HomeostasisTheory} that $1/\a$ should lie to the right of $s_{\text{max}}$. In the cell size homeostasis context, this remarkable prediction has indeed been validated experimentally in homeostatic populations for various bacterial species, growth conditions and experimental modalities~\cite{2023-emergentsimplicity,2024-mboc,2024-AnnuRevEmSim}.


{\bf \emph{Emergence of catastrophic events.}} We now turn our attention to the non-intuitive properties conferred to state variable distributions within the stochasticity-induced breakdown regime by their scale-invariant tails. In turn, we also highlight the downstream consequences for risk management. To start with, we highlight  a visually arresting trend that is evident in the nature of trajectories of the state variable, as shown in Fig.~\ref{fig-trajectories}. In this figure, we take $\Pi$ to be the same Beta distribution as in Fig.~\ref{fig-alphafig}(a) and also consider the same three values of $\a$ as used in Fig.~\ref{fig-tails}: $\a = 0.45$ is located within the homeostasis regime while $\a=0.89$ and $0.985$ are in the stochasticity-induced breakdown regime. For each of these values of $\a$, nine random trajectories of the state variable $a$ are numerically generated from Eq.~\eqref{eq-stochasticmap} and presented in next-vs-current value plots. The value of $a$ are mean-rescaled by the homeostatic mean $\le\la a\ri\ra_{\infty}= \b/(1-\a)$ for easy comparison across different $\a$ values. It is clear that when $\a$ is in the regime of homeostasis Fig~\ref{fig-trajectories}(a) (same value of $\a$ as in Fig~\ref{fig-tails}(b)), the trajectories are tightly confined about the mean (gray dot) and limited roughly by the range over which $\Pi(s)$ is nonzero (dashed rectangle; recall from Eq.~\eqref{eq-perfecthomeostasis} that at $\a=0$ the mean-rescaled state variable distribution is exactly $\Pi$). Any excursions outside these limits are highly suppressed, their probabilities of occurrence falling exponentially or faster with size (defining characteristic of a `light-tailed' distribution). However, in the heavy-tailed regime ($\a = 0.89, 0.985$), the probability of occurrence of a large deviation falls off slowly, as a power law. Thus,  intermittent giant excursions of $a$ without any obvious scale of reference (a consequence of the scale-invariant or scale-free nature of the power law tails) can be observed within the modest sample sizes shown in Figs.~\ref{fig-trajectories}(b) and (c). These catastrophes are popularly known as `black swan' events \cite{taleb2020statistical}. As $\a$ approaches the complete breakdown regime, extremely large excursions become more probable (observe how the axis ranges increase from $\a=0.89$, Fig.~\ref{fig-trajectories}(b), to $\a = 0.985$, Fig.~\ref{fig-trajectories}(c)).

\begin{figure}[h]
\begin{center}
\resizebox{0.45\textwidth}{!}{\includegraphics{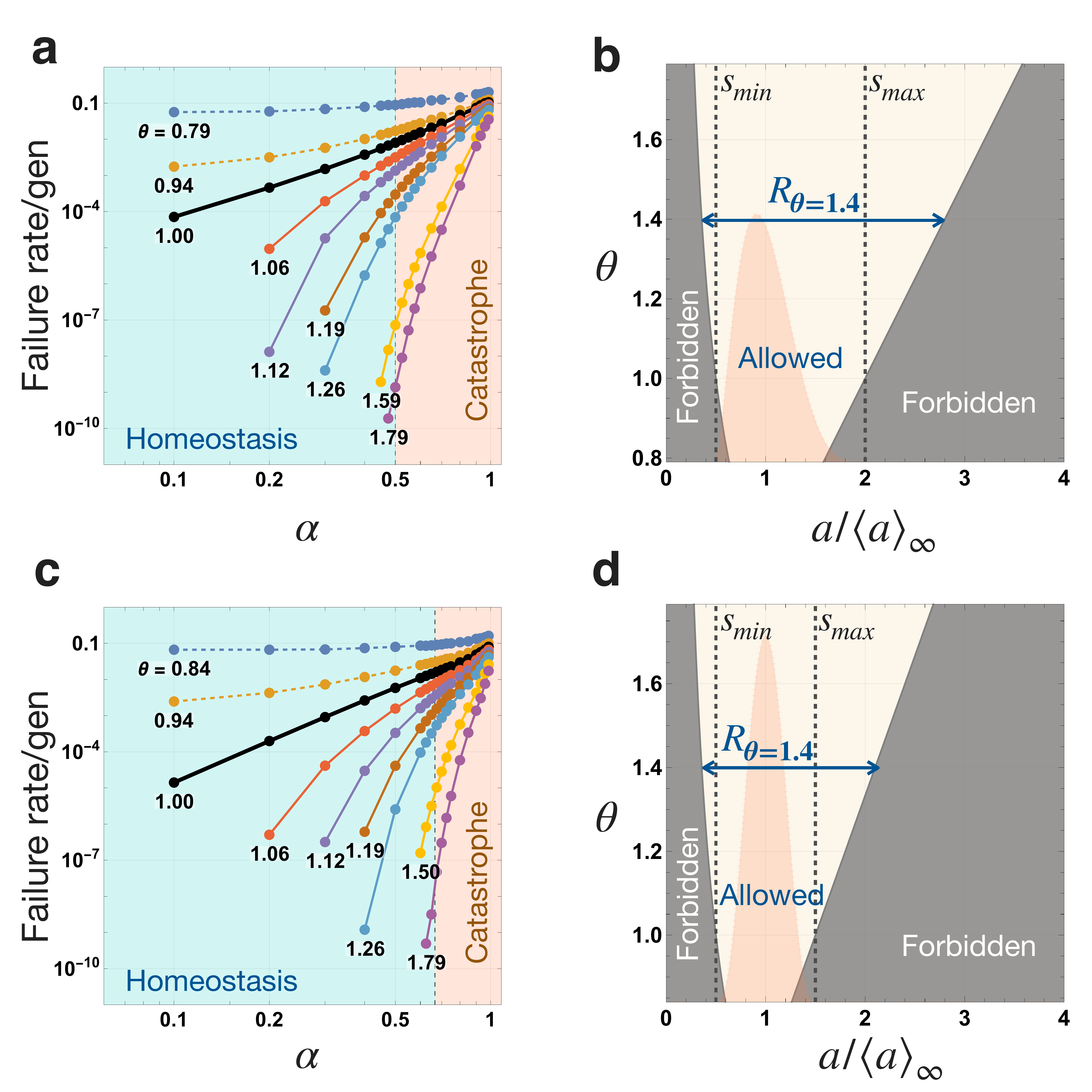}}
\caption{\textcolor{\mycolor}{\textbf{Extremal event-driven rate of failure increases sharply in the catastrophic regime.} Figs.~(a, c) show the $\a$ and $\th$-dependencies of the failure rate, using log-log axes and corresponding to the $\th$-tuned acceptable/allowed windows shown in Figs.~(b, d) respectively. In Figs.~(a, c), discs indicate numerically-obtained data points and lines are provided as visual guides; corresponding $\th$ values are indicated next to the curves. The respective $\Pi(s)$ distributions used are those shown in Figs.~\ref{fig-alphafig}(a) and \ref{fig-alphafigB}(a) (most notably, $s_{\text{max}}$ changes from $2$ to $1.5$ and so the homeostasis-catastrophe boundary shifting from $\a_{\infty} = 1/s_{\text{max}} = 0.5$ to $0.67$). For convenience of visualization, these are also shown in arbitrary units in Figs.~(b, d). For a given value of $\th$, Figs.~(b, d) allow reading off the acceptable region $R_{\th} = (s_{min}/\th, s_{max} \th)$, within which the mean-rescaled value of $a$ is allowed to vary ($R_{\th}$ for $\th=1.4$ is explicitly indicated as an example). $\th < 1$ values correspond to the region accessed by $a$, when starting from its mean value, within one generation (thus, not relevant when designing homeostatic systems). Curves corresponding to $\th < 1$ are dashed in Figs.~(a, c), while the borderline cases ($\th=1$) are thick and black. As $a$ evolves according to Eq.~\eqref{eq-stochasticmap}, it can stumble into the forbidden regions outside $R_{\th}$, indicated as the shaded portions in Figs.~(b, d), leading to failure. The failure rate is the exponential generational decay rate of the size of the portion of the population that has not visited the forbidden region(s) even once. Figs.~(a, c) show that the failure rates in the catastrophe region are relatively high and insensitive to $\th$, but are strongly suppressed for $\th>1$ values (realistic choices for designing homeostatic systems) in the homeostasis regime. Comparing Figs.~(a, c), we see that this resulting fan-like family of curves shifts with the homeostasis-catastrophe separation boundary.}}
\label{fig-failurerate}
\end{center}
\end{figure}

\textcolor{\mycolor}{{\bf \emph{Extremal event-driven rate of failure increases sharply in the catastrophic regime.}} We have just seen in a qualitative way that in the catastrophe regime, intermittent large extremal excursions occur more frequently. We now quantify this behavior inspired by living cells. Within such systems, where maintaining homeostasis is a matter of survival, excursions beyond acceptable ranges of excursion are costly and may even be fatal, leading to failure over some time period. Hard limits already exist in certain scenarios: bacterial cells cannot be smaller than a certain size to accommodate  critical life machinery, while becoming too large is metabolically unsustainable because of diffusion limitations~\cite{2015-milo,2011-dill}.  To quantify the propensity toward such failures, we designated an allowed range of mean-rescaled system variable values, $R_{\th}=(s_{min}/\th, s_{max} \th)$,  parametrized by a parameter $\th$ (Figs.~\ref{fig-failurerate}b, d). Exceeding this range during the course of its random walk causes an individual system to fail, hence the region outside $R_{\th}$ is forbidden. $R_{\th}$ has the following property: for $\th \leq 1$, there is a finite probability that an individual starting from the longterm mean value, $\le\la a\ri\ra_{\infty}$, will enter the forbidden region within a single generation, while for $\th>1$ more than one generation is necessary for this to occur. Thus, $\th>1$ values are more relevant when considering homeostatic systems that are designed to last for many generations. We will be interested in finite (i.e., not large) values of $\th$ since we are considering systems that need to constrain large deviations from homeostasis. From numerical simulations of Eq.~\eqref{eq-stochasticmap} starting from a fixed initial value, we observed that after initial condition-dependent transients, the subset of the population that has never entered the forbidden region decreased in number exponentially with each passing generation. Thus, there is a well-defined exponential failure rate that is a function of $\a$ and $\th$. This is obtained from our simulations and shown in Figs.~\ref{fig-failurerate}(a) and \ref{fig-failurerate}(c), which correspond to the $\Pi(s)$ distributions shown in Figs.~\ref{fig-alphafig}(a) and \ref{fig-alphafigB}(a) (most notably, $s_{\text{max}}$ changes from $2$ to $1.5$ and so the homeostasis-catastrophe boundary shifting from $\a_{\infty} = 1/s_{\text{max}} = 0.5$ to $0.67$). Using log-log axes, Figs.~\ref{fig-failurerate}(a) and \ref{fig-failurerate}(c) display a clear distinction between the $\th < 1$ (dashed) and $\th > 1$ curves: they are concave and convex upward, respectively, with the borderline $\th=1$ case (thick, black) being neither. This creates a characteristic fan-shaped family of curves. Near the $\a\gtrsim 1$ end where complete breakdown occurs, the failure rates are high and comparatively insensitive to $\th$. However,  as $\a$ decreases and crosses over from the catastrophe into the homeostasis regime, the $\th>1$ failure rates drop off steeply, with the steepness increasing strongly with $\th$ (regimes are labelled and indicated by backgound shading as in Figs.~\ref{fig-alphafig}(b) and \ref{fig-alphafigB}(b), inter-regime boundary marked by dashed line in Figs.~\ref{fig-failurerate}(a, c); note the logarithmic axes). Thus, as the allowed region $R_{\th}$ broadens, in the homeostasis regime the propensity of failure becomes rapidly insignificant, when compared with the insensitivity in the catastrophe regime.  Finally, as the boundary between the catastrophe and homeostasis regimes shifts between Figs.~\ref{fig-failurerate}(a) and \ref{fig-failurerate}(c), both the location of the neck of the fan shape, as well as the pattern of steep dropoff in failure rates for the significant $\th >1$ region, also shift with the homeostasis-catastrophe boundary. Thus, systems that require avoidance of extremal event-driven failures also need to remain sufficiently far away from the catastrophe regime, inside the homeostasis regime, such that the failure rate is acceptably insignificant.} 

{\bf \emph{Outlier-driven extremal behavior.}} Consider when a random sample of $N$ values drawn from a distribution totals to a sum that is extremely large compared to the expected value of $N$ times the distribution mean. When the distribution is a common light-tailed one such as a Gaussian or an Exponential, an analysis of such extremal behavior shows that the most likely method for the large sum to arise is `conspiratorial'~\cite{2013-catastrophe,2022-catastrophe,2023-catastrophe}. In other words, most of the $N$ variables are larger than the mean by moderate amounts and their cumulative yields the large excess. In contrast, for heavy tailed distributions such extremal behaviors typically involve just one outlier (a `catastrophe', as viewed from the point of view of traditional contexts where heavy tail analysis is used, such as in finance~\cite{2003-finance-book}) which accounts for the excess. 

To formulate this concept mathematically, we consider independent random variables $a_{1}, a_{2} \ldots a_{N}$, all identically varying according to the probability distribution under question. The `catastrophe' principle~\cite{2013-catastrophe,2022-catastrophe,2023-catastrophe} obeyed by heavy-tailed distributions states that the distributions of their sum, and their maximum, exhibit the same tail. Thus, for heavy-tailed distributions, the $N$-fold catastrophe measure
\begin{align}\label{eq-catastropheC}
\mc{C}_N(A) &= \frac{P(\max(a_{1} , a_{2} , \ldots , a_{N})> A)}{P(a_{1} + a_{2} + \ldots + a_{N}> A)} \to 1 \;\nn\\
&\qquad\qquad\qquad\qquad\qquad \text{ as } \; A\to \infty.
\end{align}

Since the event $\max(a_{1} , a_{2} , \ldots , a_{N})> A$ is a subset of the event $a_{1} + a_{2} + \ldots + a_{N}> A$, we always have $\mc{C}_N\leq 1$. The fact that $\mc{C}_N$ becomes $1$ indicates that any exceptional behavior of the collective is almost always the result of a single outlier. This outlier-dominated behavior stands in stark contrast with that of random variables obeying lighter-tailed distributions such as the Gaussian and Exponential distributions, where the ratio above becomes zero as $A\to \infty$ (a `conspiracy' principle). 

\begin{figure}[t]
\begin{center}
\resizebox{0.49\textwidth}{!}{\includegraphics{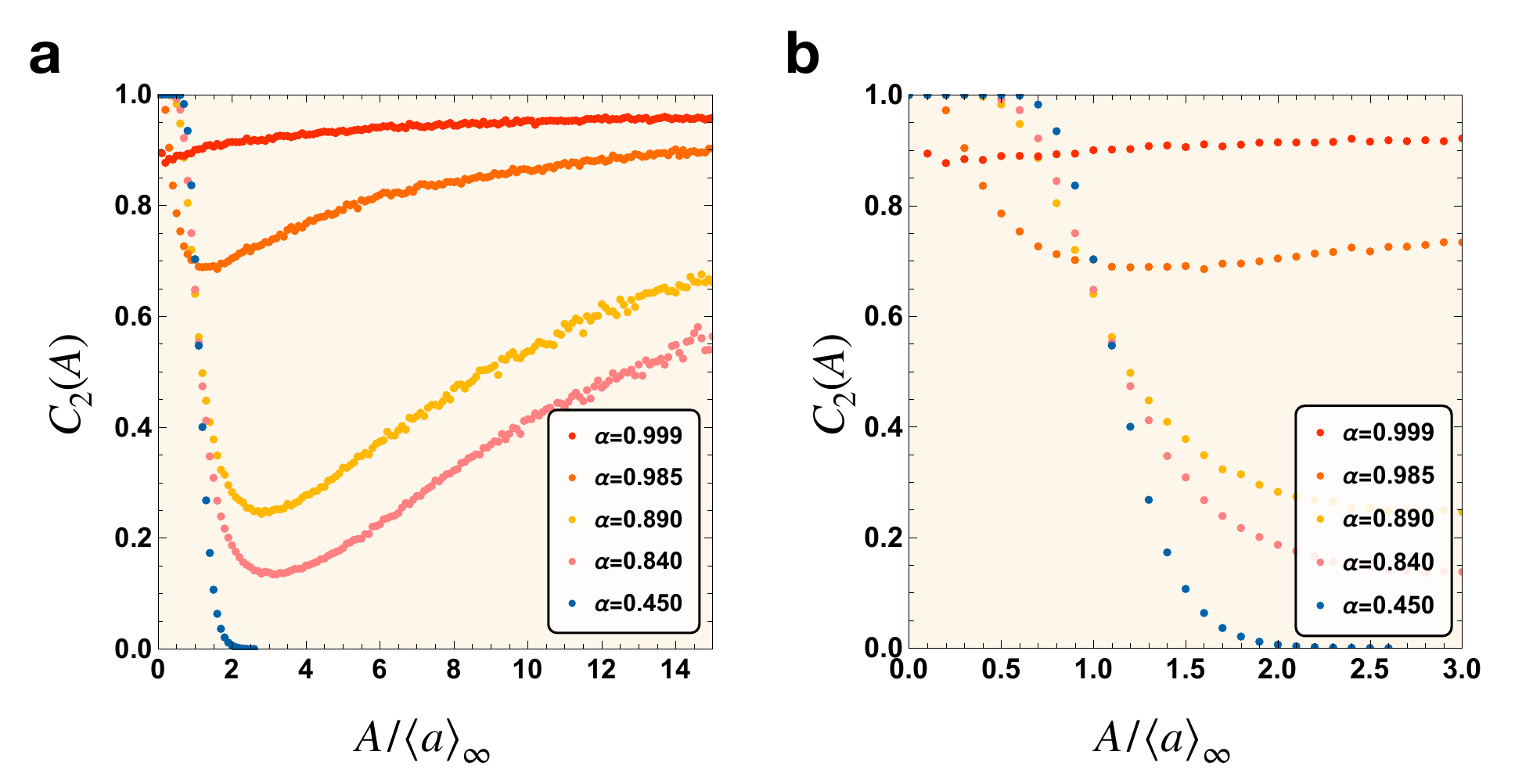}}
\caption{\textbf{Outlier-driven extremal population behavior: Catastrophic versus conspiratorial.} (a) shows the numerically calculated $2$-fold catastrophe measure of the longterm distribution (arising from Eq.~\eqref{eq-stochasticmap}), as defined in Eq.~\eqref{eq-catastropheC} with $N=2$, as a function of the threshold of extremal behavior, $A$, rescaled by the mean of the state variable, $\le\la a\ri\ra_{\infty}$, for different values of $\a$. These simulations were performed using the Beta function form for $\Pi(s)$ as shown in Fig.~\ref{fig-alphafig}(a), when $\a_{\infty}=0.5$ divides the Homeostasis regime from the Stochasticity-induced breakdown regime. Thus, $\a=0.45$ lies in the homeostasis regime (Fig.~\ref{fig-alphafig}(b)) and for this case $C_{2}$ can be seen to become zero at large $A$, showing that the homeostatic distribution satisfies the `conspiracy' principle. The remaining values of $\a$ used are located in the breakdown regime (Fig.~\ref{fig-alphafig}(b))and at large $A$, $C_{2}(A)$ can be seen to rise towards unity, showing that a single outlier is responsible for the extremal behavior of the sum. Thus, these distributions follow the `catastrophe' principle, indicating the presence of heavy tails. (b) shows a zoomed-in portion of (a) for \emph{small} values of mean-rescaled $A$, showing that $C_{2}$ jumps back up to $1$ here and the size of this region where it is $1$ progressively decreases as $\a$ goes up. This occurs because the mean-rescaled lower bound, $a_{\text{min}}$, of the state of the state variable distribution decreases monotonically with $\a$ (Eq.~\eqref{eq-amaxmin}).}
\label{fig-ratioplot}
\end{center}
\end{figure}
\begin{figure*}[t]
\begin{center}
\resizebox{0.99\textwidth}{!}{\includegraphics{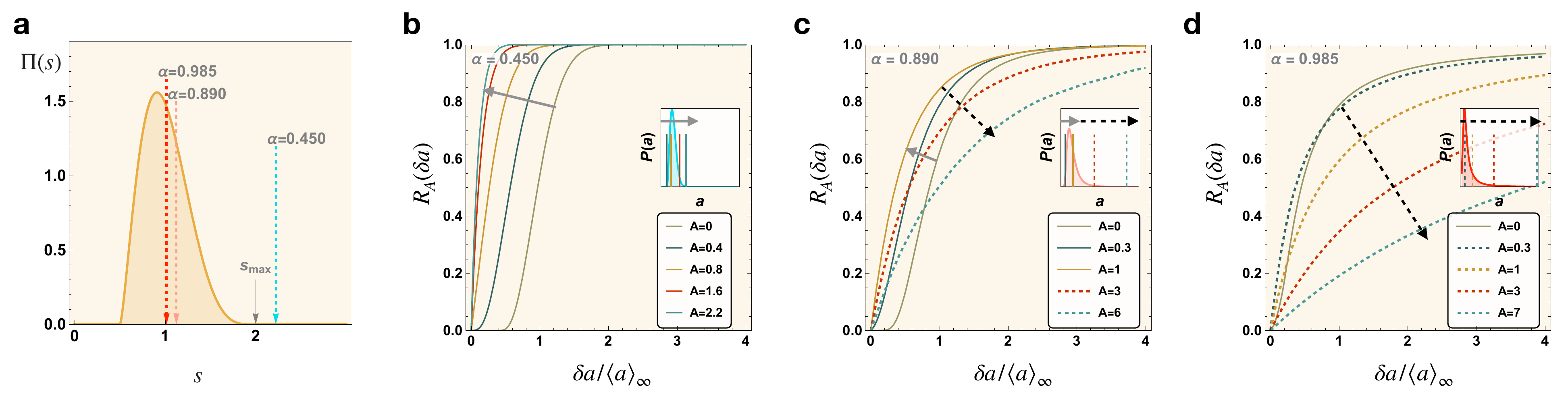}}
\caption{\textbf{Reversal of monotonicity of the conditional exceedance distributions in the stochasticity induced onset of catastrophe regime.} (a) The stretched Beta distribution form used for $\Pi(s)$ in our simulations (Fig.~\ref{fig-alphafig}(a) has more details). The locations of $1/\a$ for the three $\a$ values used in (b-2) are indicated by arrows. (b-d) Numerically generated conditional exceedance distributions (Eq.~\eqref{eq-exceedance}), $R_{A}(\d a)$, corresponding to the longterm cumulative distribution of the conditional exceedance, $\d a = a - A$, for values of $a$ larger than the threshold $A$  (both mean-rescaled by $\le\la a\ri\ra_{\infty}$). When $A$ is low enough and lies well within the conventional part of the longterm distribution of $a$ (inset), the curves rise up with increasing $A$ (solid arrows). However, when $A$ and $a$ are inside the heavy tail, this ordering with $A$ is reversed (dashed arrows). The insets show the longterm distributions of $a$: the locations of $A$-values used are shown and the conventional and heavy-tailed regions indicated by solid/dashed arrows. Since $\a=0.45$ in (b) lies inside the homeostatic regime ($1/\a$ is outside to the right of the $\Pi(s)$ distribution in (a); also see Fig.~\ref{fig-alphafig}(b)), there is only the conventional rising of the curves with $A$. For $\a=0.89$ (Fig.~(c)) inside the breakdown regime, both conventional and heavy tail behaviors can be seen. $\a=0.985$ (Fig.~(d)) approaches the complete breakdown regime and so mostly displays heavy-tail behavior.}
\label{fig-exceedance}
\end{center}
\end{figure*}

In Fig~\ref{fig-ratioplot}, we demonstrate that in the stochasticity-induced homeostasis breakdown regime the distribution of $a$ indeed obeys the catastrophe principle, i.e., outlier trajectories dominate any extremal behavior of the collective. We have numerically plotted $\mc{C}_2(A)$ for the mean-rescaled longterm distribution for the case $N=2$, for $\a$-values encompassing both homeostatic and breakdown regimes. (By iteration, one can show that demonstrating catastrophic behavior in the $N=2$ case guarantees that such behavior will occur for all $N$~\cite{1964-chistyakov}.) In the homeostatic regime, cutoffs exist on both upper and lower ends of the longterm distribution:
\begin{align}\label{eq-amaxmin}
\frac{a_{\text{m}} }{\le\la a\ri\ra_{\infty}\!\!\!}&= \frac{1-\a}{1/s_{\text{m}} - \a}, \;\; \text{m = max/min}.
\end{align}
Thus, in the homeostatic regime, $\mc{C}_N$ has to go to zero for large $A$ since the numerator in Eq.~\eqref{eq-catastropheC} becomes zero before the denominator does, as $A$ is increased. This is clearly true in the blue curve in Fig~\ref{fig-ratioplot} corresponding to the homeostatic regime and indicates the unimportance of outliers there. $a_{\text{max}}$ becomes undefined as $\a$ crosses $\a_{\infty} = 1/s_{\text{max}}$ into the heavy-tail breakdown regime, where the `catastrophe' principle comes to play. In this regime, $\a>0.5$ in Fig~\ref{fig-ratioplot}, $\mc{C}_2$ is initially less than $1$ as it behaves like a light-tailed variable, but as the power-law heavy tail starts exerting influence it picks up and goes towards $1$ as expected for outlier-dominated `catastrophic' behavior. This catastrophe-dominated region of $A$ extends to lower and lower values (of $A$) as $\a$ increases deeper into the homeostatic breakdown regime.

As an aside, in Fig~\ref{fig-ratioplot} there is a region near $A=0$ where $\mc{C}_2$ is $1$, whose size shrinks to zero as $a\to 1$. This arises because of the existence of a lower cutoff $a_{\text{min}}$ (Eq.~\eqref{eq-amaxmin}), below which the longterm probability distribution vanishes. Thus, $\mc{C}_N(A)=1$ for $A<a_{\text{min}}$ since both numerator and denominator in Eq.~\eqref{eq-catastropheC} are identically $1$. This region shrinks to zero on the mean-rescaled plot as $\a$ increases to $1$ because the mean-rescaled $a_{\min}$ monotonically decreases to zero in the interval $0<\a<1$, as can be seen from Eq.~\eqref{eq-amaxmin} (also using $s_{\text{min}}<1$).

\textcolor{\mycolor}{Equivalent quantifications of the catastrophe principle, such as path-dependent measurement without waiting for steady state, is an area of active research~\cite{2024-chen}.}

{\bf \emph{Reverse monotonicity of the conditional exceedance distribution.}} A final demonstration of outlier-driven behavior in the stochasticity-induced breakdown regime can be seen by looking at `filtered' population with state variable values greater than some large value $A$ in the tail of the distribution. What is the nature of distribution of the conditional exceedances~\cite{1974-exceedance,2013-exceedance}, $\d a = a - A$, the amounts by which these values exceed $A$? Our intuition, based on light-tailed distributions, is that the higher $A$ is, the more squished towards the origin will be this conditional exceedance distribution, and the lesser its mean. When the state variable denotes waiting time, as in queuing theory where the conditional exceedance distribution is called the residual life distribution, this intuition comes from the experience that if one has already waited for a longer time, the remaining wait time will be decreased. This expectation is however reversed for heavy tailed distributions: larger the threshold $A$, larger is the probable excess value that will be observed! (In terms of waiting times this translates to the non-intuitive experience: the longer one has waited, the longer one should expect to wait further!)

The conditional exceedance distribution is conveniently characterized by its cumulative distribution function as follows:
\begin{align}\label{eq-exceedance}
R_{A}(\d a) &= P(a<A+\d a| a>A) \nn\\
&= \frac{\int_{A}^{A+\d a} da P_{\infty}(a)}{\int_{A}^{\infty} da P_{\infty}(a)}.
\end{align}
Following the discussion above, we expect for light-tailed distributions that the cumulative function $R_{A}(\d a)$ will rise to $1$ faster for larger $A$, and so the cumulant curves for larger $A$ will lie above those for smaller $A$ in the $R(A,\d a)$ vs.\ $\d a$ plots. For heavy tailed distributions, this ordering should be reversed when $A$ is inside the heavy tail region. \textcolor{\mycolor}{In fact, in the power law regime, when $A$ is well inside the power tail, the exceedance can be explicitly calculated: $R_{A}(\d a) = 1 - (1+\d a/A)^{-k^{*}(\a)}$, where $-k^{*}(\a)$ is the Pareto exponent of the longterm size distribution calculated in Eq.~\eqref{eq-paretoexponent}.} Fig.~\ref{fig-exceedance}(b) shows that the conditional exceedance distribution follows the expected light-tailed hierarchy set by the threshold $A$ in the homeostatic regime. In the regime of stochasticity-induced breakdown of homeostasis, however, the heavy tails induce a reversal in this hierarchy (Figs.~\ref{fig-exceedance}(c), (d)). When the state variable is cell size, this phenomenon can be studied by filtering a synthetic population `designed' to be in the heavy tail regime through filters of varying sizes and observing the residue of large cells left behind. The mean size of residual individuals in such a population will increase faster than the filter pore size used to filter them.

\subsection*{Discussion}

\emph{Risk management strategies.} The results above provide a multi-faceted characterization of  the non-intuitive scale-free extremal behavior exhibited by the longtime distribution in the stochasticity-induced breakdown regime of Eq.~\eqref{eq-stochasticmap}. We now consider the potential `economics' associated with such phenomena: the potential costs versus the benefits, in contexts where extremal behavior is costly, such as untenably large sized organisms~\cite{2023-emergentsimplicity,2024-mboc,2024-AnnuRevEmSim,2023-HomeostasisTheory,2024-architecture}, devastating earthquakes~\cite{2023-earthquake}, or volatile markets~\cite{2003-finance-book}. Specifically, what system design principles can help avoid costly catastrophes, while still relaxing the rigidity of control to prevent associated costs from becoming prohibitive. While we highlight the cell size homeostasis problem below, our conclusions can be presumably be applied to other significant contexts, e.g., social and economic scenarios~\cite{1999-zipf,2003-finance-book} which incorporate processes governed by Eq.~\eqref{eq-stochasticmap} (equivalently, Eq.~\eqref{eq-kestenmap}) and where the parameter $\a$ may be controlled through, say, rules governing the marketplace and trading behaviors. Equivalently, aspects of $\Pi(s)$ may be controllable, reshaping the function $k^{*}(\a)$ in Fig.~\ref{fig-alphafig} and allowing us to achieve the constraints discussed below through an alternate route.

Focusing on control through varying $\a$ in Eq.~\eqref{eq-stochasticmap}, it seems reasonable to assume some benefit of relaxing perfect size control, i.e., having a nonzero large $\a$. For instance, the emergent scaling law, Eq.~\eqref{eq-scalinglaw}, naturally leads to cell size control through direct size control only over a cell volume reporter species~\cite{2024-architecture}, presumably leading to usage of less resources (a \emph{benefit}) compared when all components of the cell are monitored to implement perfect control. This reporter-mediated measurement however leads to less control over the cell size (the \emph{cost}), reflected in a nonzero value for $\a$~\cite{2024-architecture}. Clearly, the complete breakdown region with no longterm distribution, $\a>\a_{0}$ (Fig.~\ref{fig-alphafig}(b)), needs to be avoided since all sizes run away to infinity (thus, there is no normalizable longterm distribution) and result in the maximum cost possible. Further details of system design depend on the nature of this cost-benefit tradeoff, which we discuss in very general terms.

We will assume that the benefit function is finite and after an initial jump after some finite nonzero value of $\a$, quickly saturate or increases very slowly, since using our mechanistic insight above~\cite{2024-architecture} this benefit mainly arises due to changing the nature of size control (direct to indirect) that is responsible. In contrast, the cost of growing too large presumably increases rapidly as a polynomial or exponential in size, saturating if at all at some large value. \textcolor{\mycolor}{The section titled ``Extremal event-driven rate of failure increases sharply in the catastrophic regime'' and the accompanying Fig.~\ref{fig-failurerate} provide a possible method for motivating this cost function. An extreme scenario is when there is a truly forbidden region for some value of $\th$, straying into which the cell expires. From the corresponding failure rate in Fig.~\ref{fig-failurerate}, we can deduce that as $\a$ is tuned from the homeostasis into the catastrophe regime, the cells live for much fewer generations, produce much fewer offspring and the population rate of growth is drastically reduced. This corresponds to a rapid increase in the cost of increasing $\a$ into the catastrophe regime. Details of the context in which the cost function is calculated will determine the actual $\a$-dependence; however it is reasonable to assume that the sharply increasing nature of the function will be preserved. Further consequences will be dictated by the nature of this sharp increase.}

When the cost function is exponential in size, \textcolor{\mycolor}{its expectation value over generations} diverges when the size distribution has a power law tail. In this scenario, the entire breakdown region needs to also be avoided, i.e., $\a>\a_{\infty}$ is forbidden. If the cost function goes as a power law, $\sim a^{c}$ say, then some flexibility may be allowed. The expected cost, $\sim \le\la a^{c}\ri\ra_{\infty}$, increases rapidly with $\a$ inside the breakdown regime and becomes infinite when $\a$ crosses a threshold, $\a_{c}$, where the power law tail has the exponent $-(c+1)$. This occurs when $k^{*}(\a_{c}) = c$, where $k^{*}(\a)$ is \textcolor{\mycolor}{the function} defined implicitly in Eq.~\eqref{eq-paretoexponent} (also, Fig.~\ref{fig-alphafig}(b)). Minimizing the total cost will thus yield a point $\a<\a_{c}$, which lies in the homeostatic regime or well before $\a_{c}$ inside the stochasticity-induced breakdown regime in Fig.~\ref{fig-alphafig}(b), depending on details. \textcolor{\mycolor}{In both cases above, the cost-benefit analysis suggests why the catastrophe regime is strategically unfavorable when system preservation is desired.}

As previously mentioned, experiments performed at different growth conditions with different bacterial species and across different experimental modalities~\cite{2023-emergentsimplicity,2023-HomeostasisTheory,2024-mboc,2024-AnnuRevEmSim}, have been found to yield values of $\a$ that always lie inside the homeostatic regime and near the transition to breakdown. Experimentally, this is the observation that $1/\a$ lies just outside to the right of the experimentally measured $\Pi(s)$. From the foregoing discussion, it then appears that the cost associated with catastrophically large cell sizes is an exponentially growing function or a polynomially increasing function with a large exponent (as seen in Fig.~\ref{fig-alphafig}, if $1/\a$ lies in the experimentally undetectable tail, then $\a$ is near $\a_{k}$ values with very large $k$).

\subsection*{Concluding remarks}
We have previously \textcolor{\mycolor}{motivated, using an explicitly worked-out mechanistic model~\cite{2024-architecture},} the robust architectural underpinnings of the observed emergent simplicities in the stochastic intergenerational homeostasis of individual bacterial cells, including the scaling law Eq.~\eqref{eq-scalinglaw}~\cite{2023-emergentsimplicity}. We have also argued that such emergent simplicities~\cite{2014-PNAS,2023-emergentsimplicity} serve as constraints that deconstrain~\cite{Kirschner1998,Doyle2011,2024-AnnuRevEmSim}: the diversity of their system-specific implementations, say in different microorganisms or growth conditions, contrasts sharply with the universality of their behaviors as characterized by emergent simplicities~\cite{2023-emergentsimplicity,2014-PNAS}. In turn, these behaviors are derived from commonalities in the underlying architecture~\cite{2024-architecture,2014-PRL,2024-AnnuRevEmSim}. By constraining or limiting variations that may break the core functional architecture, these constraints that deconstrain facilitate evolvability. Thus excursions can still occur without catastrophic loss of core functionality~\cite{Kirschner1998,Doyle2011,2002-doyle}.  Within this perspective, perhaps the utility of accessing the onset of catastrophe regime becomes apparent under stressful or inclement conditions, when `black swan' excursions such as in Fig.~\ref{fig-trajectories} may provide a route to survival, despite the catastrophic risks. The clustering of observed values of $\a$ in homeostatic cells near the boundary which characterizes the onset of catastrophe~\cite{2023-HomeostasisTheory,2024-mboc,2024-AnnuRevEmSim} reflects the proximity to the exploratory frontier even when conditions are favorable.

Heavy-tailed distributions are conventionally thought of as being generated through stochastic processes that effectively already incorporate similar heavy-tailed distributions into their prescriptions (L\'evy flights, infinite node Poisson model, etc.) \cite{MANDELBROT20031,mandelbrot1982fractal,resnick2007heavy}. Remarkably, we see that the stochastic process considered in this work, Eq.~\eqref{eq-stochasticmap}, which describes homeostasis in experimentally observed systems~\cite{2023-emergentsimplicity,2024-mboc,2024-AnnuRevEmSim,2023-HomeostasisTheory,2024-architecture}, also generates heavy tailed distributions while maintaining two simple attributes: (i) Its stochasticity arises from a single well-behaved stochastic variable $s$, and (ii) it allows one-dimensional tuning all the way from well-behaved homeostasis, through a stochasticity-induced breakdown regime described by heavy-tailed distributions with predictable and, in principle, any Pareto exponent value, to a compete breakdown regime bereft of any notion of stability (Fig.~\ref{fig-alphafig}).\\

Eq.~\eqref{eq-stochasticmap}  represents a minimal model of a stochastic process which is tunable across the complete spectrum of stable, heavy-tailed or completely unstable behaviors, and furthermore is analytically tractable. Eq.~\eqref{eq-stochasticmap} is also directly related to Kesten's process, which has been speculated to be relevant for a plethora of real world complex systems, \textcolor{\mycolor}{including earthquakes~\cite{2023-earthquake}, stochastic optimization algorithms~\cite{1999-algorithms}, economic and financial markets~\cite{2003-finance-book,2023-finance}, socioeconomic behaviors~\cite{1999-zipf,2024-socioeconomic}, and social media usage~\cite{2023-socialmedia}.} How the insights from this work can inform studies of  onset of catastrophes and risk management in these more complex systems remains to be seen.

\begin{acknowledgments}
\emph{Acknowledgements}: R.R.B. and S.I.-B. gratefully acknowledge Purdue University Startup funds and the Purdue Research Foundation for supporting the collaboration and the research. S.I.-B. thanks the Purdue College of Science Dean's Special Fund and the Showalter Trust for partial support. K.J. and S.I.-B. acknowledge support from the Ross-Lynn Fellowship award and the Bilsland Dissertation Fellowship award. R.R.B. and S.I.-B. thank Mogens Jensen for helpful discussions.

\emph{Author Contributions}: RRB and SI-B conceived of and designed the research; RRB and SI-B derived the theory results and performed numerical simulations; KJ provided the argument for monotonicity of $\a_{p}$ stated between Eqs.~\eqref{eq-definealphap} and \eqref{eq-alphalimits}; RRB, CSW and SI-B wrote the paper; all authors discussed the results.
\end{acknowledgments}

\clearpage

\onecolumngrid
\pagebreak
\begin{center}
{\bf\Large Supplemental Figures}
\end{center}
\setcounter{secnumdepth}{3}  
\setcounter{equation}{0}
\setcounter{figure}{0}
\renewcommand{\theequation}{S-\arabic{equation}}
\renewcommand{\thefigure}{S\arabic{figure}}
\renewcommand\figurename{Supplementary Figure}
\renewcommand\tablename{Supplementary Table}
\newcommand\Scite[1]{[S\citealp{#1}]}
\makeatletter \renewcommand\@biblabel[1]{[S#1]} \makeatother

\section*{A second example of $\Pi(s)$}

\begin{figure*}[h]
\begin{center}
\resizebox{0.95\textwidth}{!}{\includegraphics{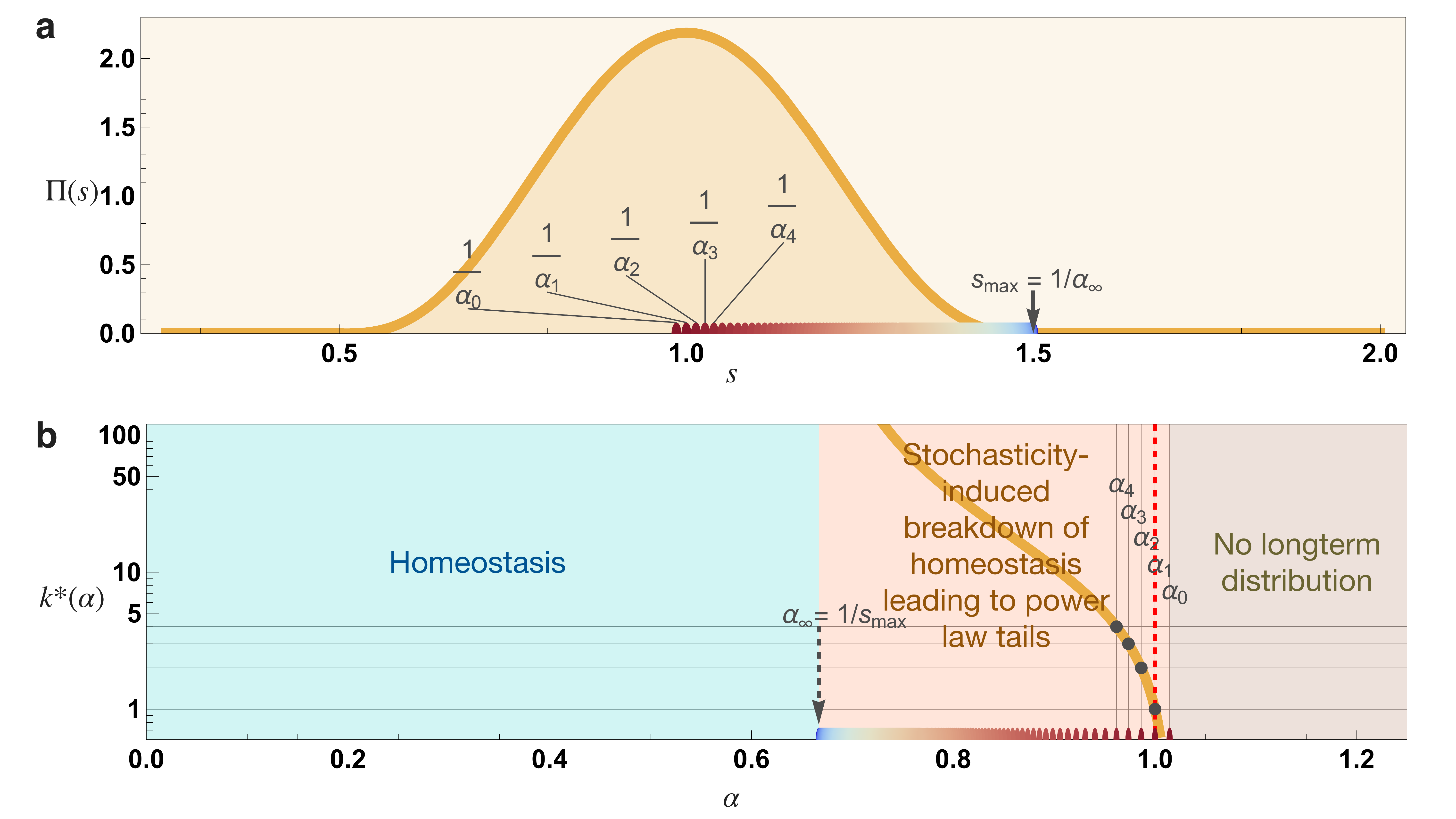}}
\caption{\textcolor{\mycolor}{\textbf{Stochasticity governs specification of distinct regimes corresponding to: homeostasis, onset of catastrophe, and system breakdown.} (a) Shows a second example $\Pi(s)$, with narrower range of support and a smaller $s_{\max}$ compared with that shown in Fig.~\ref{fig-alphafig}(a). It is a Beta distribution with parameters $(4,4)$ in standard notation, stretched and translated to lie between $s_{\text{min}}=0.5$ and $s_{\text{max}}=1.5$. The mean is $1$ as required. (b) Shows three different dynamical regimes that occur as $\a$ is varied in Eq.~\eqref{eq-stochasticmap}. Compare this with Fig.~\ref{fig-alphafig}(b); the same quantities and regimes are indicated here but these are now calculated using the new distribution shown in (a).}}
\label{fig-alphafigB}
\end{center}
\end{figure*}

\clearpage

\pagebreak

\section*{Convergence to steady state}

\begin{figure*}[h]
\begin{center}
\resizebox{0.45\textwidth}{!}{\includegraphics{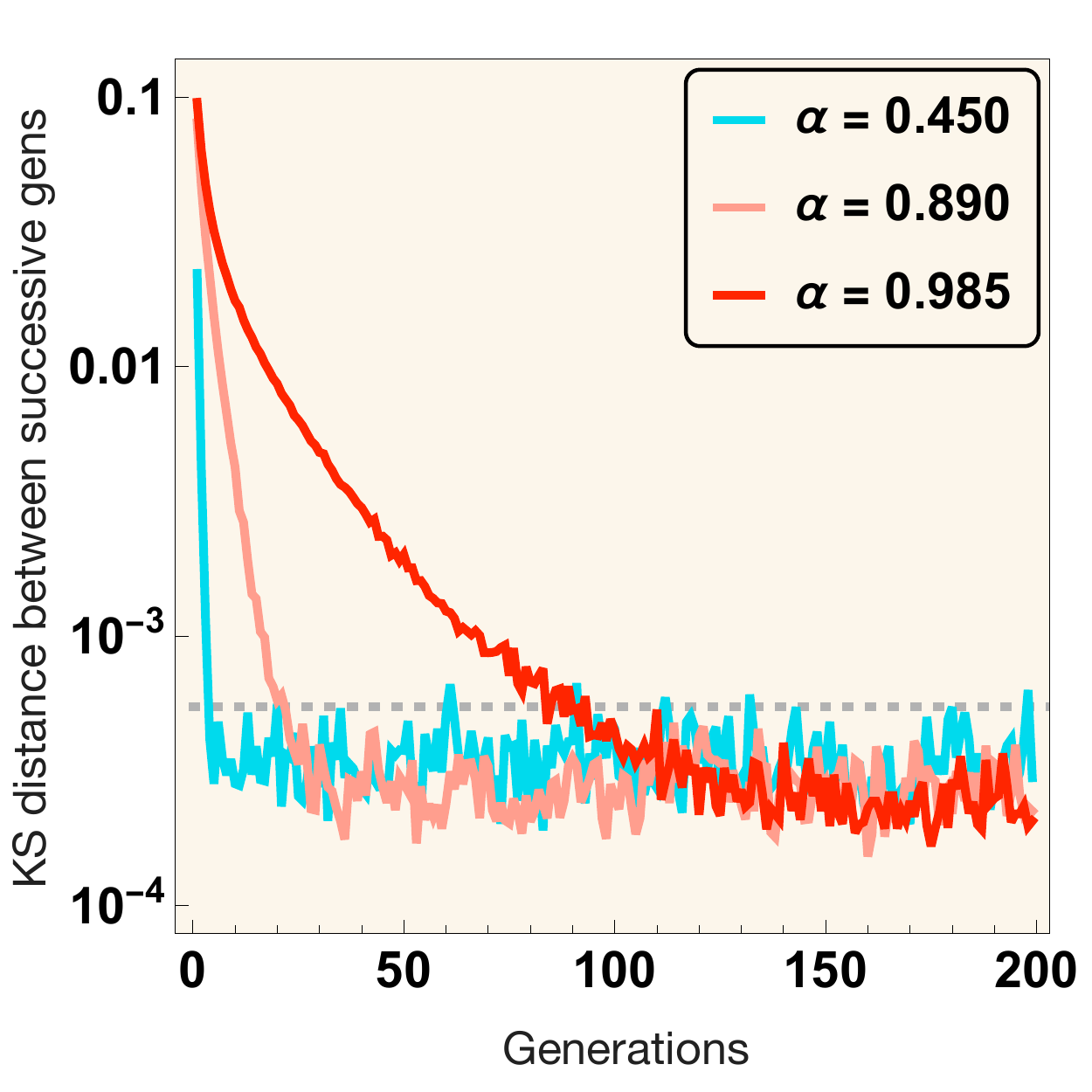}}
\caption{\textcolor{\mycolor}{\textbf{Convergence to a final steady state distribution.} This figure plots the Kolmogorov-Smirnov (KS) distance between the $a$-distributions at successive generations as, for different values of $\a$, $10$ million identical systems evolve intergenerationally according to Eq.~\eqref{eq-stochasticmap}, starting from the initial condition $a_{0}=1$ (in mean-rescaled units) and with $\Pi(s)$ given by Fig.~\ref{fig-alphafig}(a). The rapid exponential-like decrease to a sample size-limited minimum value demonstrates the convergence to a steady state distribution (note that the vertical axis uses the logarithmic scale). As $\a$ increases into the catastrophe regime ($\a > \a_{\infty} = 0.5$), this convergence slows down as it takes more generations to fill in the long heavy tails of the steady state distribution. These three curves correspond to the $\a$-values used to plot the steady state power law tails in Fig.~\ref{fig-tails}. The gray dashed line indicates the $95$ percentile limit of the appropriately scaled large sample limit of the Kolmogorov distribution, indicating the sample size-limited saturation value of the metric being plotted. (Technical note: For increasing $\a$, successive generations become more correlated and the saturation distribution becomes progressively narrower than the Kolmogorov distribution, as can be seen above.)}}
\label{fig-convergence}
\end{center}
\end{figure*}



\begin{thebibliography}{62}%
\makeatletter
\providecommand \@ifxundefined [1]{%
 \@ifx{#1\undefined}
}%
\providecommand \@ifnum [1]{%
 \ifnum #1\expandafter \@firstoftwo
 \else \expandafter \@secondoftwo
 \fi
}%
\providecommand \@ifx [1]{%
 \ifx #1\expandafter \@firstoftwo
 \else \expandafter \@secondoftwo
 \fi
}%
\providecommand \natexlab [1]{#1}%
\providecommand \enquote  [1]{``#1''}%
\providecommand \bibnamefont  [1]{#1}%
\providecommand \bibfnamefont [1]{#1}%
\providecommand \citenamefont [1]{#1}%
\providecommand \href@noop [0]{\@secondoftwo}%
\providecommand \href [0]{\begingroup \@sanitize@url \@href}%
\providecommand \@href[1]{\@@startlink{#1}\@@href}%
\providecommand \@@href[1]{\endgroup#1\@@endlink}%
\providecommand \@sanitize@url [0]{\catcode `\\12\catcode `\$12\catcode
  `\&12\catcode `\#12\catcode `\^12\catcode `\_12\catcode `\%12\relax}%
\providecommand \@@startlink[1]{}%
\providecommand \@@endlink[0]{}%
\providecommand \url  [0]{\begingroup\@sanitize@url \@url }%
\providecommand \@url [1]{\endgroup\@href {#1}{\urlprefix }}%
\providecommand \urlprefix  [0]{URL }%
\providecommand \Eprint [0]{\href }%
\providecommand \doibase [0]{https://doi.org/}%
\providecommand \selectlanguage [0]{\@gobble}%
\providecommand \bibinfo  [0]{\@secondoftwo}%
\providecommand \bibfield  [0]{\@secondoftwo}%
\providecommand \translation [1]{[#1]}%
\providecommand \BibitemOpen [0]{}%
\providecommand \bibitemStop [0]{}%
\providecommand \bibitemNoStop [0]{.\EOS\space}%
\providecommand \EOS [0]{\spacefactor3000\relax}%
\providecommand \BibitemShut  [1]{\csname bibitem#1\endcsname}%
\let\auto@bib@innerbib\@empty
\bibitem [{\citenamefont {Miller}(2016)}]{2016-miller}%
  \BibitemOpen
  \bibfield  {author} {\bibinfo {author} {\bibfnamefont {J.~H.}\ \bibnamefont
  {Miller}},\ }\href@noop {} {\emph {\bibinfo {title} {A crude look at the
  whole}}}\ (\bibinfo  {publisher} {Basic Books},\ \bibinfo {address} {New
  York},\ \bibinfo {year} {2016})\BibitemShut {NoStop}%
\bibitem [{\citenamefont {Wiener}(1961)}]{1961-wiener}%
  \BibitemOpen
  \bibfield  {author} {\bibinfo {author} {\bibfnamefont {N.}~\bibnamefont
  {Wiener}},\ }\href@noop {} {\emph {\bibinfo {title} {Cybernetics}}},\
  \bibinfo {edition} {2nd}\ ed.\ (\bibinfo  {publisher} {The MIT Press},\
  \bibinfo {address} {Cambridge, MA},\ \bibinfo {year} {1961})\BibitemShut
  {NoStop}%
\bibitem [{\citenamefont {Ramsay}\ and\ \citenamefont
  {Woods}(2016)}]{2016-ramsay}%
  \BibitemOpen
  \bibfield  {author} {\bibinfo {author} {\bibfnamefont {D.~S.}\ \bibnamefont
  {Ramsay}}\ and\ \bibinfo {author} {\bibfnamefont {S.~C.}\ \bibnamefont
  {Woods}},\ }\bibfield  {title} {\bibinfo {title} {Physiological regulation:
  How it really works},\ }\href {https://doi.org/10.1016/j.cmet.2016.08.004}
  {\bibfield  {journal} {\bibinfo  {journal} {Cell Metabolism}\ }\textbf
  {\bibinfo {volume} {24}},\ \bibinfo {pages} {361} (\bibinfo {year}
  {2016})}\BibitemShut {NoStop}%
\bibitem [{\citenamefont {Kotas}\ and\ \citenamefont
  {Medzhitov}(2015)}]{2015-kotas}%
  \BibitemOpen
  \bibfield  {author} {\bibinfo {author} {\bibfnamefont {M.~E.}\ \bibnamefont
  {Kotas}}\ and\ \bibinfo {author} {\bibfnamefont {R.}~\bibnamefont
  {Medzhitov}},\ }\bibfield  {title} {\bibinfo {title} {Homeostasis,
  inflammation, and disease susceptibility},\ }\href
  {https://doi.org/10.1016/j.cell.2015.02.010} {\bibfield  {journal} {\bibinfo
  {journal} {Cell}\ }\textbf {\bibinfo {volume} {160}},\ \bibinfo {pages} {816}
  (\bibinfo {year} {2015})}\BibitemShut {NoStop}%
\bibitem [{\citenamefont {Wiener}(1951)}]{1951-wiener}%
  \BibitemOpen
  \bibfield  {author} {\bibinfo {author} {\bibfnamefont {N.}~\bibnamefont
  {Wiener}},\ }\bibfield  {title} {\bibinfo {title} {Homeostasis in the
  individual and society},\ }\href
  {https://doi.org/10.1016/0016-0032(51)90897-6} {\bibfield  {journal}
  {\bibinfo  {journal} {Journal of the Franklin Institute}\ }\textbf {\bibinfo
  {volume} {251}},\ \bibinfo {pages} {65} (\bibinfo {year} {1951})}\BibitemShut
  {NoStop}%
\bibitem [{\citenamefont {Yi}\ \emph {et~al.}(2000)\citenamefont {Yi},
  \citenamefont {Huang}, \citenamefont {Simon},\ and\ \citenamefont
  {Doyle}}]{2000-yi}%
  \BibitemOpen
  \bibfield  {author} {\bibinfo {author} {\bibfnamefont {T.-M.}\ \bibnamefont
  {Yi}}, \bibinfo {author} {\bibfnamefont {Y.}~\bibnamefont {Huang}}, \bibinfo
  {author} {\bibfnamefont {M.~I.}\ \bibnamefont {Simon}},\ and\ \bibinfo
  {author} {\bibfnamefont {J.}~\bibnamefont {Doyle}},\ }\bibfield  {title}
  {\bibinfo {title} {Robust perfect adaptation in bacterial chemotaxis through
  integral feedback control},\ }\href {https://doi.org/10.1073/pnas.97.9.4649}
  {\bibfield  {journal} {\bibinfo  {journal} {Proceedings of the National
  Academy of Sciences}\ }\textbf {\bibinfo {volume} {97}},\ \bibinfo {pages}
  {4649} (\bibinfo {year} {2000})}\BibitemShut {NoStop}%
\bibitem [{\citenamefont {Karin}\ \emph {et~al.}(2016)\citenamefont {Karin},
  \citenamefont {Swisa}, \citenamefont {Glaser}, \citenamefont {Dor},\ and\
  \citenamefont {Alon}}]{2016-karin}%
  \BibitemOpen
  \bibfield  {author} {\bibinfo {author} {\bibfnamefont {O.}~\bibnamefont
  {Karin}}, \bibinfo {author} {\bibfnamefont {A.}~\bibnamefont {Swisa}},
  \bibinfo {author} {\bibfnamefont {B.}~\bibnamefont {Glaser}}, \bibinfo
  {author} {\bibfnamefont {Y.}~\bibnamefont {Dor}},\ and\ \bibinfo {author}
  {\bibfnamefont {U.}~\bibnamefont {Alon}},\ }\bibfield  {title} {\bibinfo
  {title} {Dynamical compensation in physiological circuits},\ }\href
  {https://doi.org/10.15252/msb.20167216} {\bibfield  {journal} {\bibinfo
  {journal} {Molecular Systems Biology}\ }\textbf {\bibinfo {volume} {12}},\
  \bibinfo {pages} {886} (\bibinfo {year} {2016})}\BibitemShut {NoStop}%
\bibitem [{\citenamefont {Topp}\ \emph {et~al.}(2000)\citenamefont {Topp},
  \citenamefont {Promislow}, \citenamefont {deVries}, \citenamefont {Miura},\
  and\ \citenamefont {Finegood}}]{2000-topp}%
  \BibitemOpen
  \bibfield  {author} {\bibinfo {author} {\bibfnamefont {B.}~\bibnamefont
  {Topp}}, \bibinfo {author} {\bibfnamefont {K.}~\bibnamefont {Promislow}},
  \bibinfo {author} {\bibfnamefont {G.}~\bibnamefont {deVries}}, \bibinfo
  {author} {\bibfnamefont {R.~M.}\ \bibnamefont {Miura}},\ and\ \bibinfo
  {author} {\bibfnamefont {D.~T.}\ \bibnamefont {Finegood}},\ }\bibfield
  {title} {\bibinfo {title} {A model of beta-cell mass, insulin, and glucose
  kinetics: Pathways to diabetes},\ }\href
  {https://doi.org/10.1006/jtbi.2000.2150} {\bibfield  {journal} {\bibinfo
  {journal} {Journal of Theoretical Biology}\ }\textbf {\bibinfo {volume}
  {206}},\ \bibinfo {pages} {605} (\bibinfo {year} {2000})}\BibitemShut
  {NoStop}%
\bibitem [{\citenamefont {El-Samad}\ \emph {et~al.}(2002)\citenamefont
  {El-Samad}, \citenamefont {Goff},\ and\ \citenamefont
  {Khammash}}]{2002-el-samad}%
  \BibitemOpen
  \bibfield  {author} {\bibinfo {author} {\bibfnamefont {H.}~\bibnamefont
  {El-Samad}}, \bibinfo {author} {\bibfnamefont {J.~P.}\ \bibnamefont {Goff}},\
  and\ \bibinfo {author} {\bibfnamefont {M.}~\bibnamefont {Khammash}},\
  }\bibfield  {title} {\bibinfo {title} {Calcium homeostasis and parturient
  hypocalcemia: An integral feedback perspective},\ }\href
  {https://doi.org/10.1006/jtbi.2001.2422} {\bibfield  {journal} {\bibinfo
  {journal} {Journal of Theoretical Biology}\ }\textbf {\bibinfo {volume}
  {214}},\ \bibinfo {pages} {17} (\bibinfo {year} {2002})}\BibitemShut
  {NoStop}%
\bibitem [{\citenamefont {Frere}\ and\ \citenamefont
  {Slutsky}(2018)}]{2018-frere}%
  \BibitemOpen
  \bibfield  {author} {\bibinfo {author} {\bibfnamefont {S.}~\bibnamefont
  {Frere}}\ and\ \bibinfo {author} {\bibfnamefont {I.}~\bibnamefont
  {Slutsky}},\ }\bibfield  {title} {\bibinfo {title} {Alzheimer's disease: From
  firing instability to homeostasis network collapse},\ }\href
  {https://doi.org/10.1016/j.neuron.2017.11.028} {\bibfield  {journal}
  {\bibinfo  {journal} {Neuron}\ }\textbf {\bibinfo {volume} {97}},\ \bibinfo
  {pages} {32} (\bibinfo {year} {2018})}\BibitemShut {NoStop}%
\bibitem [{\citenamefont {Korem~Kohanim}\ \emph {et~al.}(2020)\citenamefont
  {Korem~Kohanim}, \citenamefont {Tendler}, \citenamefont {Mayo}, \citenamefont
  {Friedman},\ and\ \citenamefont {Alon}}]{2020-korem-kohanim}%
  \BibitemOpen
  \bibfield  {author} {\bibinfo {author} {\bibfnamefont {Y.}~\bibnamefont
  {Korem~Kohanim}}, \bibinfo {author} {\bibfnamefont {A.}~\bibnamefont
  {Tendler}}, \bibinfo {author} {\bibfnamefont {A.}~\bibnamefont {Mayo}},
  \bibinfo {author} {\bibfnamefont {N.}~\bibnamefont {Friedman}},\ and\
  \bibinfo {author} {\bibfnamefont {U.}~\bibnamefont {Alon}},\ }\bibfield
  {title} {\bibinfo {title} {Endocrine autoimmune disease as a fragility of
  immune surveillance against hypersecreting mutants},\ }\href@noop {}
  {\bibfield  {journal} {\bibinfo  {journal} {Immunity}\ }\textbf {\bibinfo
  {volume} {52}},\ \bibinfo {pages} {872} (\bibinfo {year} {2020})}\BibitemShut
  {NoStop}%
\bibitem [{\citenamefont {Murphy}\ and\ \citenamefont
  {Jones}(2018)}]{2018-McGrath}%
  \BibitemOpen
  \bibfield  {author} {\bibinfo {author} {\bibfnamefont {K.~G.}\ \bibnamefont
  {Murphy}}\ and\ \bibinfo {author} {\bibfnamefont {N.~S.}\ \bibnamefont
  {Jones}},\ }\bibfield  {title} {\bibinfo {title} {Quantitative approaches to
  energy and glucose homeostasis: machine learning and modelling for precision
  understanding and prediction},\ }\href
  {https://doi.org/10.1098/rsif.2017.0736} {\bibfield  {journal} {\bibinfo
  {journal} {J. R. Soc. Interface.}\ }\textbf {\bibinfo {volume} {15}},\
  \bibinfo {pages} {20170736} (\bibinfo {year} {2018})}\BibitemShut {NoStop}%
\bibitem [{\citenamefont {Nijhout}\ \emph {et~al.}(2014)\citenamefont
  {Nijhout}, \citenamefont {Best},\ and\ \citenamefont {Reed}}]{2014-nijhout}%
  \BibitemOpen
  \bibfield  {author} {\bibinfo {author} {\bibfnamefont {H.~F.}\ \bibnamefont
  {Nijhout}}, \bibinfo {author} {\bibfnamefont {J.}~\bibnamefont {Best}},\ and\
  \bibinfo {author} {\bibfnamefont {M.~C.}\ \bibnamefont {Reed}},\ }\bibfield
  {title} {\bibinfo {title} {Escape from homeostasis},\ }\href
  {https://doi.org/10.1016/j.mbs.2014.08.015} {\bibfield  {journal} {\bibinfo
  {journal} {Mathematical Biosciences}\ }\textbf {\bibinfo {volume} {257}},\
  \bibinfo {pages} {104} (\bibinfo {year} {2014})}\BibitemShut {NoStop}%
\bibitem [{\citenamefont {Joshi}\ \emph {et~al.}(2024)\citenamefont {Joshi},
  \citenamefont {Wright}, \citenamefont {Biswas},\ and\ \citenamefont
  {Iyer-Biswas}}]{2024-architecture}%
  \BibitemOpen
  \bibfield  {author} {\bibinfo {author} {\bibfnamefont {K.}~\bibnamefont
  {Joshi}}, \bibinfo {author} {\bibfnamefont {C.~S.}\ \bibnamefont {Wright}},
  \bibinfo {author} {\bibfnamefont {R.~R.}\ \bibnamefont {Biswas}},\ and\
  \bibinfo {author} {\bibfnamefont {S.}~\bibnamefont {Iyer-Biswas}},\
  }\bibfield  {title} {\bibinfo {title} {Architectural underpinnings of
  stochastic intergenerational homeostasis},\ }\href
  {https://doi.org/10.1103/PhysRevE.110.024405} {\bibfield  {journal} {\bibinfo
   {journal} {Phys. Rev. E}\ }\textbf {\bibinfo {volume} {110}},\ \bibinfo
  {pages} {024405} (\bibinfo {year} {2024})}\BibitemShut {NoStop}%
\bibitem [{\citenamefont {Iyer-Biswas}\ \emph
  {et~al.}(2014{\natexlab{a}})\citenamefont {Iyer-Biswas}, \citenamefont
  {Wright}, \citenamefont {Henry}, \citenamefont {Lo}, \citenamefont {Burov},
  \citenamefont {Lin}, \citenamefont {Crooks}, \citenamefont {Crosson},
  \citenamefont {Dinner},\ and\ \citenamefont {Scherer}}]{2014-PNAS}%
  \BibitemOpen
  \bibfield  {author} {\bibinfo {author} {\bibfnamefont {S.}~\bibnamefont
  {Iyer-Biswas}}, \bibinfo {author} {\bibfnamefont {C.~S.}\ \bibnamefont
  {Wright}}, \bibinfo {author} {\bibfnamefont {J.~T.}\ \bibnamefont {Henry}},
  \bibinfo {author} {\bibfnamefont {K.}~\bibnamefont {Lo}}, \bibinfo {author}
  {\bibfnamefont {S.}~\bibnamefont {Burov}}, \bibinfo {author} {\bibfnamefont
  {Y.}~\bibnamefont {Lin}}, \bibinfo {author} {\bibfnamefont {G.~E.}\
  \bibnamefont {Crooks}}, \bibinfo {author} {\bibfnamefont {S.}~\bibnamefont
  {Crosson}}, \bibinfo {author} {\bibfnamefont {A.~R.}\ \bibnamefont
  {Dinner}},\ and\ \bibinfo {author} {\bibfnamefont {N.~F.}\ \bibnamefont
  {Scherer}},\ }\bibfield  {title} {\bibinfo {title} {Scaling laws governing
  stochastic growth and division of single bacterial cells},\ }\href
  {https://doi.org/10.1073/pnas.1403232111} {\bibfield  {journal} {\bibinfo
  {journal} {Proc. Natl. Acad. Sci. U.S.A.}\ }\textbf {\bibinfo {volume}
  {111}},\ \bibinfo {pages} {15912} (\bibinfo {year}
  {2014}{\natexlab{a}})}\BibitemShut {NoStop}%
\bibitem [{\citenamefont {Joshi}\ \emph
  {et~al.}(2023{\natexlab{a}})\citenamefont {Joshi}, \citenamefont {Wright},
  \citenamefont {Ziegler}, \citenamefont {Spiers}, \citenamefont {Crosser},
  \citenamefont {Eschker}, \citenamefont {Biswas},\ and\ \citenamefont
  {Iyer-Biswas}}]{2023-emergentsimplicity}%
  \BibitemOpen
  \bibfield  {author} {\bibinfo {author} {\bibfnamefont {K.}~\bibnamefont
  {Joshi}}, \bibinfo {author} {\bibfnamefont {C.~S.}\ \bibnamefont {Wright}},
  \bibinfo {author} {\bibfnamefont {K.~F.}\ \bibnamefont {Ziegler}}, \bibinfo
  {author} {\bibfnamefont {E.~M.}\ \bibnamefont {Spiers}}, \bibinfo {author}
  {\bibfnamefont {J.~T.}\ \bibnamefont {Crosser}}, \bibinfo {author}
  {\bibfnamefont {S.}~\bibnamefont {Eschker}}, \bibinfo {author} {\bibfnamefont
  {R.~R.}\ \bibnamefont {Biswas}},\ and\ \bibinfo {author} {\bibfnamefont
  {S.}~\bibnamefont {Iyer-Biswas}},\ }\bibfield  {title} {\bibinfo {title}
  {Emergent simplicities in stochastic intergenerational homeostasis},\
  }\bibfield  {journal} {\bibinfo  {journal} {bioRxiv:2023.01.18.524627}\
  }\href {https://doi.org/10.1101/2023.01.18.524627}
  {10.1101/2023.01.18.524627} (\bibinfo {year}
  {2023}{\natexlab{a}})\BibitemShut {NoStop}%
\bibitem [{\citenamefont {Ziegler}\ \emph {et~al.}(2024)\citenamefont
  {Ziegler}, \citenamefont {Joshi}, \citenamefont {Wright}, \citenamefont
  {Roy}, \citenamefont {Caruso}, \citenamefont {Biswas},\ and\ \citenamefont
  {Iyer-Biswas}}]{2024-mboc}%
  \BibitemOpen
  \bibfield  {author} {\bibinfo {author} {\bibfnamefont {K.~F.}\ \bibnamefont
  {Ziegler}}, \bibinfo {author} {\bibfnamefont {K.}~\bibnamefont {Joshi}},
  \bibinfo {author} {\bibfnamefont {C.~S.}\ \bibnamefont {Wright}}, \bibinfo
  {author} {\bibfnamefont {S.}~\bibnamefont {Roy}}, \bibinfo {author}
  {\bibfnamefont {W.}~\bibnamefont {Caruso}}, \bibinfo {author} {\bibfnamefont
  {R.~R.}\ \bibnamefont {Biswas}},\ and\ \bibinfo {author} {\bibfnamefont
  {S.}~\bibnamefont {Iyer-Biswas}},\ }\bibfield  {title} {\bibinfo {title}
  {Scaling of stochastic growth and division dynamics: A comparative study of
  individual rod-shaped cells in the mother machine and schemostat platforms},\
  }\href {https://doi.org/10.1091/mbc.E23-11-0452} {\bibfield  {journal}
  {\bibinfo  {journal} {Molecular Biology of the Cell}\ }\textbf {\bibinfo
  {volume} {35:ar78}},\ \bibinfo {pages} {1} (\bibinfo {year}
  {2024})}\BibitemShut {NoStop}%
\bibitem [{\citenamefont {Wright}\ \emph {et~al.}(2024)\citenamefont {Wright},
  \citenamefont {Joshi}, \citenamefont {Biswas},\ and\ \citenamefont
  {Iyer-Biswas}}]{2024-AnnuRevEmSim}%
  \BibitemOpen
  \bibfield  {author} {\bibinfo {author} {\bibfnamefont {C.~S.}\ \bibnamefont
  {Wright}}, \bibinfo {author} {\bibfnamefont {K.}~\bibnamefont {Joshi}},
  \bibinfo {author} {\bibfnamefont {R.~R.}\ \bibnamefont {Biswas}},\ and\
  \bibinfo {author} {\bibfnamefont {S.}~\bibnamefont {Iyer-Biswas}},\
  }\bibfield  {title} {\bibinfo {title} {Emergent simplicities in the living
  histories of individual cells},\ }\Eprint {https://arxiv.org/abs/2404.01682}
  {arXiv:2404.01682}  (\bibinfo {year} {2024}),\ \bibinfo {note} {(In press,
  Annual Review of Condensed Matter Physics)}\BibitemShut {NoStop}%
\bibitem [{\citenamefont {Joshi}\ \emph
  {et~al.}(2023{\natexlab{b}})\citenamefont {Joshi}, \citenamefont {Biswas},\
  and\ \citenamefont {Iyer-Biswas}}]{2023-HomeostasisTheory}%
  \BibitemOpen
  \bibfield  {author} {\bibinfo {author} {\bibfnamefont {K.}~\bibnamefont
  {Joshi}}, \bibinfo {author} {\bibfnamefont {R.~R.}\ \bibnamefont {Biswas}},\
  and\ \bibinfo {author} {\bibfnamefont {S.}~\bibnamefont {Iyer-Biswas}},\
  }\bibfield  {title} {\bibinfo {title} {Intergenerational scaling law
  determines the precision kinematics of stochastic individual-cell-size
  homeostasis},\ }\href@noop {} {\bibfield  {journal} {\bibinfo  {journal}
  {bioRxiv:2023.01.20.525000}\ } (\bibinfo {year}
  {2023}{\natexlab{b}})}\BibitemShut {NoStop}%
\bibitem [{\citenamefont {Jun}\ and\ \citenamefont
  {Taheri-Araghi}(2015)}]{Jun2015}%
  \BibitemOpen
  \bibfield  {author} {\bibinfo {author} {\bibfnamefont {S.}~\bibnamefont
  {Jun}}\ and\ \bibinfo {author} {\bibfnamefont {S.}~\bibnamefont
  {Taheri-Araghi}},\ }\bibfield  {title} {\bibinfo {title} {Cell-size
  maintenance: Universal strategy revealed},\ }\href
  {https://doi.org/10.1016/j.tim.2014.12.001} {\bibfield  {journal} {\bibinfo
  {journal} {Trends in Microbiology}\ }\textbf {\bibinfo {volume} {23}},\
  \bibinfo {pages} {4} (\bibinfo {year} {2015})}\BibitemShut {NoStop}%
\bibitem [{\citenamefont {Willis}\ and\ \citenamefont
  {Huang}(2017)}]{Willis2017}%
  \BibitemOpen
  \bibfield  {author} {\bibinfo {author} {\bibfnamefont {L.}~\bibnamefont
  {Willis}}\ and\ \bibinfo {author} {\bibfnamefont {K.~C.}\ \bibnamefont
  {Huang}},\ }\bibfield  {title} {\bibinfo {title} {Sizing up the bacterial
  cell cycle},\ }\href {https://doi.org/10.1038/nrmicro.2017.79} {\bibfield
  {journal} {\bibinfo  {journal} {Nature Reviews Microbiology}\ }\textbf
  {\bibinfo {volume} {15}},\ \bibinfo {pages} {606} (\bibinfo {year}
  {2017})}\BibitemShut {NoStop}%
\bibitem [{\citenamefont {Modi}\ \emph {et~al.}(2017)\citenamefont {Modi},
  \citenamefont {Vargas-Garcia}, \citenamefont {Ghusinga},\ and\ \citenamefont
  {Singh}}]{Modi2017}%
  \BibitemOpen
  \bibfield  {author} {\bibinfo {author} {\bibfnamefont {S.}~\bibnamefont
  {Modi}}, \bibinfo {author} {\bibfnamefont {C.~A.}\ \bibnamefont
  {Vargas-Garcia}}, \bibinfo {author} {\bibfnamefont {K.~R.}\ \bibnamefont
  {Ghusinga}},\ and\ \bibinfo {author} {\bibfnamefont {A.}~\bibnamefont
  {Singh}},\ }\bibfield  {title} {\bibinfo {title} {Analysis of noise
  mechanisms in cell-size control},\ }\href@noop {} {\bibfield  {journal}
  {\bibinfo  {journal} {Biophys. J.}\ }\textbf {\bibinfo {volume} {112}},\
  \bibinfo {pages} {2408} (\bibinfo {year} {2017})}\BibitemShut {NoStop}%
\bibitem [{\citenamefont {Lin}\ and\ \citenamefont
  {Amir}(2017)}]{lin2017effects}%
  \BibitemOpen
  \bibfield  {author} {\bibinfo {author} {\bibfnamefont {J.}~\bibnamefont
  {Lin}}\ and\ \bibinfo {author} {\bibfnamefont {A.}~\bibnamefont {Amir}},\
  }\bibfield  {title} {\bibinfo {title} {The effects of stochasticity at the
  single-cell level and cell size control on the population growth},\
  }\href@noop {} {\bibfield  {journal} {\bibinfo  {journal} {Cell Syst.}\
  }\textbf {\bibinfo {volume} {5}},\ \bibinfo {pages} {358} (\bibinfo {year}
  {2017})}\BibitemShut {NoStop}%
\bibitem [{\citenamefont {Logsdon}\ \emph {et~al.}(2017)\citenamefont
  {Logsdon}, \citenamefont {Ho}, \citenamefont {Papavinasasundaram},
  \citenamefont {Richardson}, \citenamefont {Cokol}, \citenamefont {Sassetti},
  \citenamefont {Amir},\ and\ \citenamefont {Aldridge}}]{logsdon2017parallel}%
  \BibitemOpen
  \bibfield  {author} {\bibinfo {author} {\bibfnamefont {M.~M.}\ \bibnamefont
  {Logsdon}}, \bibinfo {author} {\bibfnamefont {P.-Y.}\ \bibnamefont {Ho}},
  \bibinfo {author} {\bibfnamefont {K.}~\bibnamefont {Papavinasasundaram}},
  \bibinfo {author} {\bibfnamefont {K.}~\bibnamefont {Richardson}}, \bibinfo
  {author} {\bibfnamefont {M.}~\bibnamefont {Cokol}}, \bibinfo {author}
  {\bibfnamefont {C.~M.}\ \bibnamefont {Sassetti}}, \bibinfo {author}
  {\bibfnamefont {A.}~\bibnamefont {Amir}},\ and\ \bibinfo {author}
  {\bibfnamefont {B.~B.}\ \bibnamefont {Aldridge}},\ }\bibfield  {title}
  {\bibinfo {title} {A parallel adder coordinates mycobacterial cell-cycle
  progression and cell-size homeostasis in the context of asymmetric growth and
  organization},\ }\href@noop {} {\bibfield  {journal} {\bibinfo  {journal}
  {Current Biology}\ }\textbf {\bibinfo {volume} {27}},\ \bibinfo {pages}
  {3367} (\bibinfo {year} {2017})}\BibitemShut {NoStop}%
\bibitem [{\citenamefont {Sauls}\ \emph {et~al.}(2016)\citenamefont {Sauls},
  \citenamefont {Li},\ and\ \citenamefont {Jun}}]{sauls2016adder}%
  \BibitemOpen
  \bibfield  {author} {\bibinfo {author} {\bibfnamefont {J.~T.}\ \bibnamefont
  {Sauls}}, \bibinfo {author} {\bibfnamefont {D.}~\bibnamefont {Li}},\ and\
  \bibinfo {author} {\bibfnamefont {S.}~\bibnamefont {Jun}},\ }\bibfield
  {title} {\bibinfo {title} {Adder and a coarse-grained approach to cell size
  homeostasis in bacteria},\ }\href@noop {} {\bibfield  {journal} {\bibinfo
  {journal} {Current opinion in cell biology}\ }\textbf {\bibinfo {volume}
  {38}},\ \bibinfo {pages} {38} (\bibinfo {year} {2016})}\BibitemShut {NoStop}%
\bibitem [{\citenamefont {Taheri-Araghi}\ \emph {et~al.}(2015)\citenamefont
  {Taheri-Araghi}, \citenamefont {Bradde}, \citenamefont {Sauls}, \citenamefont
  {Hill}, \citenamefont {Levin}, \citenamefont {Paulsson}, \citenamefont
  {Vergassola},\ and\ \citenamefont {Jun}}]{2015-JunReview}%
  \BibitemOpen
  \bibfield  {author} {\bibinfo {author} {\bibfnamefont {S.}~\bibnamefont
  {Taheri-Araghi}}, \bibinfo {author} {\bibfnamefont {S.}~\bibnamefont
  {Bradde}}, \bibinfo {author} {\bibfnamefont {J.~T.}\ \bibnamefont {Sauls}},
  \bibinfo {author} {\bibfnamefont {N.~S.}\ \bibnamefont {Hill}}, \bibinfo
  {author} {\bibfnamefont {P.~A.}\ \bibnamefont {Levin}}, \bibinfo {author}
  {\bibfnamefont {J.}~\bibnamefont {Paulsson}}, \bibinfo {author}
  {\bibfnamefont {M.}~\bibnamefont {Vergassola}},\ and\ \bibinfo {author}
  {\bibfnamefont {S.}~\bibnamefont {Jun}},\ }\bibfield  {title} {\bibinfo
  {title} {Cell-size control and homeostasis in bacteria},\ }\href
  {https://doi.org/10.1016/j.cub.2014.12.009} {\bibfield  {journal} {\bibinfo
  {journal} {Current Biology}\ }\textbf {\bibinfo {volume} {25}},\ \bibinfo
  {pages} {385} (\bibinfo {year} {2015})}\BibitemShut {NoStop}%
\bibitem [{\citenamefont {Deforet}\ \emph {et~al.}(2015)\citenamefont
  {Deforet}, \citenamefont {van Ditmarsch},\ and\ \citenamefont
  {Xavier}}]{deforet2015cell}%
  \BibitemOpen
  \bibfield  {author} {\bibinfo {author} {\bibfnamefont {M.}~\bibnamefont
  {Deforet}}, \bibinfo {author} {\bibfnamefont {D.}~\bibnamefont {van
  Ditmarsch}},\ and\ \bibinfo {author} {\bibfnamefont {J.~B.}\ \bibnamefont
  {Xavier}},\ }\bibfield  {title} {\bibinfo {title} {Cell-size homeostasis and
  the incremental rule in a bacterial pathogen},\ }\href@noop {} {\bibfield
  {journal} {\bibinfo  {journal} {Biophysical journal}\ }\textbf {\bibinfo
  {volume} {109}},\ \bibinfo {pages} {521} (\bibinfo {year}
  {2015})}\BibitemShut {NoStop}%
\bibitem [{\citenamefont {Amir}(2014)}]{amir2014cell}%
  \BibitemOpen
  \bibfield  {author} {\bibinfo {author} {\bibfnamefont {A.}~\bibnamefont
  {Amir}},\ }\bibfield  {title} {\bibinfo {title} {Cell size regulation in
  bacteria},\ }\href@noop {} {\bibfield  {journal} {\bibinfo  {journal}
  {Physical Review Letters}\ }\textbf {\bibinfo {volume} {112}},\ \bibinfo
  {pages} {208102} (\bibinfo {year} {2014})}\BibitemShut {NoStop}%
\bibitem [{\citenamefont {Spiesser}\ \emph {et~al.}(2012)\citenamefont
  {Spiesser}, \citenamefont {M{\"u}ller}, \citenamefont {Schreiber},
  \citenamefont {Krantz},\ and\ \citenamefont {Klipp}}]{Spiesser2012}%
  \BibitemOpen
  \bibfield  {author} {\bibinfo {author} {\bibfnamefont {T.~W.}\ \bibnamefont
  {Spiesser}}, \bibinfo {author} {\bibfnamefont {C.}~\bibnamefont
  {M{\"u}ller}}, \bibinfo {author} {\bibfnamefont {G.}~\bibnamefont
  {Schreiber}}, \bibinfo {author} {\bibfnamefont {M.}~\bibnamefont {Krantz}},\
  and\ \bibinfo {author} {\bibfnamefont {E.}~\bibnamefont {Klipp}},\ }\bibfield
   {title} {\bibinfo {title} {Size homeostasis can be intrinsic to growing cell
  populations and explained without size sensing or signalling},\ }\href@noop
  {} {\bibfield  {journal} {\bibinfo  {journal} {The FEBS Journal}\ }\textbf
  {\bibinfo {volume} {279}},\ \bibinfo {pages} {4213} (\bibinfo {year}
  {2012})}\BibitemShut {NoStop}%
\bibitem [{\citenamefont {Nurse}(2000)}]{2000-nurse}%
  \BibitemOpen
  \bibfield  {author} {\bibinfo {author} {\bibfnamefont {P.}~\bibnamefont
  {Nurse}},\ }\bibfield  {title} {\bibinfo {title} {A long twentieth century of
  the cell cycle and beyond},\ }\href
  {https://doi.org/10.1016/S0092-8674(00)81684-0} {\bibfield  {journal}
  {\bibinfo  {journal} {Cell}\ }\textbf {\bibinfo {volume} {100}},\ \bibinfo
  {pages} {71} (\bibinfo {year} {2000})}\BibitemShut {NoStop}%
\bibitem [{\citenamefont {Tyson}(1987)}]{1987-tyson}%
  \BibitemOpen
  \bibfield  {author} {\bibinfo {author} {\bibfnamefont {J.~J.}\ \bibnamefont
  {Tyson}},\ }\bibfield  {title} {\bibinfo {title} {Size control of cell
  division},\ }\href {https://doi.org/10.1016/S0022-5193(87)80146-7} {\bibfield
   {journal} {\bibinfo  {journal} {Journal of Theoretical Biology}\ }\textbf
  {\bibinfo {volume} {126}},\ \bibinfo {pages} {381} (\bibinfo {year}
  {1987})}\BibitemShut {NoStop}%
\bibitem [{\citenamefont {Fantes}\ and\ \citenamefont
  {Nurse}(1981)}]{1981-nurse}%
  \BibitemOpen
  \bibfield  {author} {\bibinfo {author} {\bibfnamefont {P.}~\bibnamefont
  {Fantes}}\ and\ \bibinfo {author} {\bibfnamefont {P.}~\bibnamefont {Nurse}},\
  }\href
  {https://www.cambridge.org/us/universitypress/subjects/life-sciences/cell-biology-and-developmental-biology/cell-cycle-1}
  {\emph {\bibinfo {title} {The cell cycle}}}\ (\bibinfo  {publisher}
  {Cambridge University Press},\ \bibinfo {year} {1981})\ Chap.~\bibinfo
  {chapter} {2}\BibitemShut {NoStop}%
\bibitem [{\citenamefont {Kesten}(1973)}]{1973-kesten}%
  \BibitemOpen
  \bibfield  {author} {\bibinfo {author} {\bibfnamefont {H.}~\bibnamefont
  {Kesten}},\ }\bibfield  {title} {\bibinfo {title} {Random difference
  equations and renewal theory for products of random matrices},\ }\href
  {https://doi.org/10.1007/BF02392040} {\bibfield  {journal} {\bibinfo
  {journal} {Acta Mathematica}\ }\textbf {\bibinfo {volume} {131}},\ \bibinfo
  {pages} {207} (\bibinfo {year} {1973})}\BibitemShut {NoStop}%
\bibitem [{\citenamefont {Lei}\ and\ \citenamefont
  {Sornette}(2023)}]{2023-earthquake}%
  \BibitemOpen
  \bibfield  {author} {\bibinfo {author} {\bibfnamefont {Q.}~\bibnamefont
  {Lei}}\ and\ \bibinfo {author} {\bibfnamefont {D.}~\bibnamefont {Sornette}},\
  }\bibfield  {title} {\bibinfo {title} {A stochastic dynamical model of slope
  creep and failure},\ }\href
  {https://doi.org/https://doi.org/10.1029/2022GL102587} {\bibfield  {journal}
  {\bibinfo  {journal} {Geophysical Research Letters}\ }\textbf {\bibinfo
  {volume} {50}},\ \bibinfo {pages} {e2022GL102587} (\bibinfo {year}
  {2023})}\BibitemShut {NoStop}%
\bibitem [{\citenamefont {Diaconis}\ and\ \citenamefont
  {Freedman}(1999)}]{1999-algorithms}%
  \BibitemOpen
  \bibfield  {author} {\bibinfo {author} {\bibfnamefont {P.}~\bibnamefont
  {Diaconis}}\ and\ \bibinfo {author} {\bibfnamefont {D.}~\bibnamefont
  {Freedman}},\ }\bibfield  {title} {\bibinfo {title} {Iterated random
  functions},\ }\href {https://doi.org/10.1137/S0036144598338446} {\bibfield
  {journal} {\bibinfo  {journal} {SIAM Review}\ }\textbf {\bibinfo {volume}
  {41}},\ \bibinfo {pages} {45} (\bibinfo {year} {1999})}\BibitemShut {NoStop}%
\bibitem [{\citenamefont {Finkenstadt}\ and\ \citenamefont
  {Rootz{\'e}n}(2003)}]{2003-finance-book}%
  \BibitemOpen
  \bibfield  {author} {\bibinfo {author} {\bibfnamefont {B.}~\bibnamefont
  {Finkenstadt}}\ and\ \bibinfo {author} {\bibfnamefont {H.}~\bibnamefont
  {Rootz{\'e}n}},\ }\href@noop {} {\emph {\bibinfo {title} {Extreme values in
  finance, telecommunications, and the environment}}}\ (\bibinfo  {publisher}
  {CRC Press},\ \bibinfo {year} {2003})\BibitemShut {NoStop}%
\bibitem [{\citenamefont {Francq}\ \emph {et~al.}(2023)\citenamefont {Francq},
  \citenamefont {Kandji},\ and\ \citenamefont {Zakoian}}]{2023-finance}%
  \BibitemOpen
  \bibfield  {author} {\bibinfo {author} {\bibfnamefont {C.}~\bibnamefont
  {Francq}}, \bibinfo {author} {\bibfnamefont {B.~M.}\ \bibnamefont {Kandji}},\
  and\ \bibinfo {author} {\bibfnamefont {J.-M.}\ \bibnamefont {Zakoian}},\
  }\bibfield  {title} {\bibinfo {title} {Inference on garch-midas models
  without any small-order moment},\ }\href
  {https://doi.org/10.1017/S0266466623000142} {\bibfield  {journal} {\bibinfo
  {journal} {Econometric Theory}\ ,\ \bibinfo {pages} {1}} (\bibinfo {year}
  {2023})}\BibitemShut {NoStop}%
\bibitem [{\citenamefont {Gabaix}(1999)}]{1999-zipf}%
  \BibitemOpen
  \bibfield  {author} {\bibinfo {author} {\bibfnamefont {X.}~\bibnamefont
  {Gabaix}},\ }\bibfield  {title} {\bibinfo {title} {Zipf's law for cities: An
  explanation},\ }\href {https://www.jstor.org/stable/2586883} {\bibfield
  {journal} {\bibinfo  {journal} {The Quarterly Journal of Economics}\ }\textbf
  {\bibinfo {volume} {114}},\ \bibinfo {pages} {739} (\bibinfo {year}
  {1999})}\BibitemShut {NoStop}%
\bibitem [{\citenamefont {Boppart}\ \emph {et~al.}(2024)\citenamefont
  {Boppart}, \citenamefont {Krusell},\ and\ \citenamefont
  {Olsson}}]{2024-socioeconomic}%
  \BibitemOpen
  \bibfield  {author} {\bibinfo {author} {\bibfnamefont {T.}~\bibnamefont
  {Boppart}}, \bibinfo {author} {\bibfnamefont {P.}~\bibnamefont {Krusell}},\
  and\ \bibinfo {author} {\bibfnamefont {J.}~\bibnamefont {Olsson}},\ }\href
  {https://doi.org/10.3386/w32977} {\emph {\bibinfo {title} {Who should work
  how much?}}},\ \bibinfo {type} {Working Paper}\ \bibinfo {number} {32977}\
  (\bibinfo  {institution} {National Bureau of Economic Research},\ \bibinfo
  {year} {2024})\BibitemShut {NoStop}%
\bibitem [{\citenamefont {Jiang}\ \emph {et~al.}(2023)\citenamefont {Jiang},
  \citenamefont {Yamada}, \citenamefont {Takayasu},\ and\ \citenamefont
  {Takayasu}}]{2023-socialmedia}%
  \BibitemOpen
  \bibfield  {author} {\bibinfo {author} {\bibfnamefont {J.~J.}\ \bibnamefont
  {Jiang}}, \bibinfo {author} {\bibfnamefont {K.}~\bibnamefont {Yamada}},
  \bibinfo {author} {\bibfnamefont {H.}~\bibnamefont {Takayasu}},\ and\
  \bibinfo {author} {\bibfnamefont {M.}~\bibnamefont {Takayasu}},\ }\bibfield
  {title} {\bibinfo {title} {Scale-dependent power law properties in hashtag
  usage time series of weibo},\ }\href
  {https://doi.org/10.1038/s41598-023-49572-6} {\bibfield  {journal} {\bibinfo
  {journal} {Scientific Reports}\ }\textbf {\bibinfo {volume} {13}},\ \bibinfo
  {pages} {22298} (\bibinfo {year} {2023})}\BibitemShut {NoStop}%
\bibitem [{\citenamefont {Gueneau}\ \emph {et~al.}(2023)\citenamefont
  {Gueneau}, \citenamefont {Majumdar},\ and\ \citenamefont
  {Schehr}}]{2023-gueneau}%
  \BibitemOpen
  \bibfield  {author} {\bibinfo {author} {\bibfnamefont {M.}~\bibnamefont
  {Gueneau}}, \bibinfo {author} {\bibfnamefont {S.~N.}\ \bibnamefont
  {Majumdar}},\ and\ \bibinfo {author} {\bibfnamefont {G.}~\bibnamefont
  {Schehr}},\ }\bibfield  {title} {\bibinfo {title} {Active particle in a
  harmonic trap driven by a resetting noise: an approach via kesten
  variables},\ }\href {https://doi.org/10.1088/1751-8121/ad00ef} {\bibfield
  {journal} {\bibinfo  {journal} {Journal of Physics A: Mathematical and
  Theoretical}\ }\textbf {\bibinfo {volume} {56}},\ \bibinfo {pages} {475002}
  (\bibinfo {year} {2023})}\BibitemShut {NoStop}%
\bibitem [{\citenamefont {Biroli}\ \emph {et~al.}(2024)\citenamefont {Biroli},
  \citenamefont {Kulkarni}, \citenamefont {Majumdar},\ and\ \citenamefont
  {Schehr}}]{2024-biroli}%
  \BibitemOpen
  \bibfield  {author} {\bibinfo {author} {\bibfnamefont {M.}~\bibnamefont
  {Biroli}}, \bibinfo {author} {\bibfnamefont {M.}~\bibnamefont {Kulkarni}},
  \bibinfo {author} {\bibfnamefont {S.~N.}\ \bibnamefont {Majumdar}},\ and\
  \bibinfo {author} {\bibfnamefont {G.}~\bibnamefont {Schehr}},\ }\bibfield
  {title} {\bibinfo {title} {Dynamically emergent correlations between
  particles in a switching harmonic trap},\ }\href
  {https://doi.org/10.1103/PhysRevE.109.L032106} {\bibfield  {journal}
  {\bibinfo  {journal} {Phys. Rev. E}\ }\textbf {\bibinfo {volume} {109}},\
  \bibinfo {pages} {L032106} (\bibinfo {year} {2024})}\BibitemShut {NoStop}%
\bibitem [{\citenamefont {Buraczewski}\ \emph {et~al.}(2016)\citenamefont
  {Buraczewski}, \citenamefont {Damek},\ and\ \citenamefont
  {Mikosch}}]{2016-buraczewski}%
  \BibitemOpen
  \bibfield  {author} {\bibinfo {author} {\bibfnamefont {D.}~\bibnamefont
  {Buraczewski}}, \bibinfo {author} {\bibfnamefont {E.}~\bibnamefont {Damek}},\
  and\ \bibinfo {author} {\bibfnamefont {T.}~\bibnamefont {Mikosch}},\ }\href
  {https://doi.org/10.1007/978-3-319-29679-1} {\emph {\bibinfo {title}
  {Stochastic Models with Power-Law Tails: The Equation X = AX + B}}}\
  (\bibinfo  {publisher} {Springer Cham},\ \bibinfo {year} {2016})\BibitemShut
  {NoStop}%
\bibitem [{\citenamefont {Goldie}(1991)}]{1991-goldie}%
  \BibitemOpen
  \bibfield  {author} {\bibinfo {author} {\bibfnamefont {C.~M.}\ \bibnamefont
  {Goldie}},\ }\bibfield  {title} {\bibinfo {title} {Implicit renewal theory
  and tails of solutions of random equations},\ }\href
  {https://doi.org/10.1214/aoap/1177005985} {\bibfield  {journal} {\bibinfo
  {journal} {The Annals of Applied Probability}\ }\textbf {\bibinfo {volume}
  {1}},\ \bibinfo {pages} {126 } (\bibinfo {year} {1991})}\BibitemShut
  {NoStop}%
\bibitem [{\citenamefont {Resnick}(2007)}]{resnick2007heavy}%
  \BibitemOpen
  \bibfield  {author} {\bibinfo {author} {\bibfnamefont {S.~I.}\ \bibnamefont
  {Resnick}},\ }\href@noop {} {\emph {\bibinfo {title} {Heavy-tail phenomena:
  probabilistic and statistical modeling}}}\ (\bibinfo  {publisher} {Springer
  Science \& Business Media},\ \bibinfo {year} {2007})\BibitemShut {NoStop}%
\bibitem [{\citenamefont {Mandelbrot}(2003)}]{MANDELBROT20031}%
  \BibitemOpen
  \bibfield  {author} {\bibinfo {author} {\bibfnamefont {B.~B.}\ \bibnamefont
  {Mandelbrot}},\ }\bibfield  {title} {\bibinfo {title} {Chapter 1 - heavy
  tails in finance for independent or multifractal price increments},\ }in\
  \href@noop {} {\emph {\bibinfo {booktitle} {Handbook of Heavy Tailed
  Distributions in Finance}}},\ \bibinfo {series} {Handbooks in Finance},
  Vol.~\bibinfo {volume} {1},\ \bibinfo {editor} {edited by\ \bibinfo {editor}
  {\bibfnamefont {S.~T.}\ \bibnamefont {Rachev}}}\ (\bibinfo  {publisher}
  {North-Holland},\ \bibinfo {address} {Amsterdam},\ \bibinfo {year} {2003})\
  pp.\ \bibinfo {pages} {1--34}\BibitemShut {NoStop}%
\bibitem [{\citenamefont {Mandelbrot}(1982)}]{mandelbrot1982fractal}%
  \BibitemOpen
  \bibfield  {author} {\bibinfo {author} {\bibfnamefont {B.~B.}\ \bibnamefont
  {Mandelbrot}},\ }\href@noop {} {\emph {\bibinfo {title} {The fractal geometry
  of nature}}}\ (\bibinfo  {publisher} {WH freeman New York},\ \bibinfo {year}
  {1982})\BibitemShut {NoStop}%
\bibitem [{\citenamefont {Nair}\ \emph {et~al.}(2022)\citenamefont {Nair},
  \citenamefont {Wierman},\ and\ \citenamefont {Zwart}}]{2022-catastrophe}%
  \BibitemOpen
  \bibfield  {author} {\bibinfo {author} {\bibfnamefont {J.}~\bibnamefont
  {Nair}}, \bibinfo {author} {\bibfnamefont {A.}~\bibnamefont {Wierman}},\ and\
  \bibinfo {author} {\bibfnamefont {B.}~\bibnamefont {Zwart}},\ }\href@noop {}
  {\emph {\bibinfo {title} {The fundamentals of heavy tails: Properties,
  emergence, and estimation}}}\ (\bibinfo  {publisher} {Cambridge University
  Press},\ \bibinfo {year} {2022})\BibitemShut {NoStop}%
\bibitem [{\citenamefont {Alsmeyer}\ \emph {et~al.}(2009)\citenamefont
  {Alsmeyer}, \citenamefont {Iksanov},\ and\ \citenamefont
  {R{\"o}sler}}]{2009-alsmeyer}%
  \BibitemOpen
  \bibfield  {author} {\bibinfo {author} {\bibfnamefont {G.}~\bibnamefont
  {Alsmeyer}}, \bibinfo {author} {\bibfnamefont {A.}~\bibnamefont {Iksanov}},\
  and\ \bibinfo {author} {\bibfnamefont {U.}~\bibnamefont {R{\"o}sler}},\
  }\bibfield  {title} {\bibinfo {title} {On distributional properties of
  perpetuities},\ }\href {https://doi.org/10.1007/s10959-008-0156-8} {\bibfield
   {journal} {\bibinfo  {journal} {Journal of Theoretical Probability}\
  }\textbf {\bibinfo {volume} {22}},\ \bibinfo {pages} {666} (\bibinfo {year}
  {2009})}\BibitemShut {NoStop}%
\bibitem [{\citenamefont {Taleb}(2020)}]{taleb2020statistical}%
  \BibitemOpen
  \bibfield  {author} {\bibinfo {author} {\bibfnamefont {N.~N.}\ \bibnamefont
  {Taleb}},\ }\bibfield  {title} {\bibinfo {title} {Statistical consequences of
  fat tails: Real world preasymptotics, epistemology, and applications},\
  }\href@noop {} {\bibfield  {journal} {\bibinfo  {journal} {arXiv:2001.10488}\
  } (\bibinfo {year} {2020})}\BibitemShut {NoStop}%
\bibitem [{\citenamefont {Milo}(2015)}]{2015-milo}%
  \BibitemOpen
  \bibfield  {author} {\bibinfo {author} {\bibfnamefont {R.}~\bibnamefont
  {Milo}, \bibfnamefont {R.nand~Phillips}},\ }\href
  {https://doi.org/10.1201/9780429258770} {\emph {\bibinfo {title} {Cell
  Biology by the Numbers}}}\ (\bibinfo  {publisher} {Garland Science},\
  \bibinfo {year} {2015})\BibitemShut {NoStop}%
\bibitem [{\citenamefont {Dill}\ \emph {et~al.}(2011)\citenamefont {Dill},
  \citenamefont {Ghosh},\ and\ \citenamefont {Schmit}}]{2011-dill}%
  \BibitemOpen
  \bibfield  {author} {\bibinfo {author} {\bibfnamefont {K.~A.}\ \bibnamefont
  {Dill}}, \bibinfo {author} {\bibfnamefont {K.}~\bibnamefont {Ghosh}},\ and\
  \bibinfo {author} {\bibfnamefont {J.~D.}\ \bibnamefont {Schmit}},\ }\bibfield
   {title} {\bibinfo {title} {Physical limits of cells and proteomes},\ }\href
  {https://doi.org/10.1073/pnas.1114477108} {\bibfield  {journal} {\bibinfo
  {journal} {Proceedings of the National Academy of Sciences}\ }\textbf
  {\bibinfo {volume} {108}},\ \bibinfo {pages} {17876} (\bibinfo {year}
  {2011})}\BibitemShut {NoStop}%
\bibitem [{\citenamefont {Wierman}(2013)}]{2013-catastrophe}%
  \BibitemOpen
  \bibfield  {author} {\bibinfo {author} {\bibfnamefont {A.}~\bibnamefont
  {Wierman}},\ }\href
  {https://rigorandrelevance.wordpress.com/2013/12/17/catastrophes-conspiracies-and-subexponential-distributions-part-ii/}
  {\bibinfo {title} {Catastrophes, conspiracies, and subexponential
  distributions (part ii)}} (\bibinfo {year} {2013})\BibitemShut {NoStop}%
\bibitem [{\citenamefont {Wang}\ and\ \citenamefont
  {Rhee}(2023)}]{2023-catastrophe}%
  \BibitemOpen
  \bibfield  {author} {\bibinfo {author} {\bibfnamefont {X.}~\bibnamefont
  {Wang}}\ and\ \bibinfo {author} {\bibfnamefont {C.-H.}\ \bibnamefont
  {Rhee}},\ }\bibfield  {title} {\bibinfo {title} {Importance sampling strategy
  for heavy-tailed systems with catastrophe principle},\ }in\ \href
  {https://doi.org/10.1109/WSC60868.2023.10407503} {\emph {\bibinfo {booktitle}
  {2023 Winter Simulation Conference (WSC)}}}\ (\bibinfo {year} {2023})\ pp.\
  \bibinfo {pages} {76--90}\BibitemShut {NoStop}%
\bibitem [{\citenamefont {Chistyakov}(1964)}]{1964-chistyakov}%
  \BibitemOpen
  \bibfield  {author} {\bibinfo {author} {\bibfnamefont {V.~P.}\ \bibnamefont
  {Chistyakov}},\ }\bibfield  {title} {\bibinfo {title} {A theorem on sums of
  independent positive random variables and its applications to branching
  random processes},\ }\href {https://doi.org/10.1137/1109088} {\bibfield
  {journal} {\bibinfo  {journal} {Theory of Probability \& Its Applications}\
  }\textbf {\bibinfo {volume} {9}},\ \bibinfo {pages} {640} (\bibinfo {year}
  {1964})}\BibitemShut {NoStop}%
\bibitem [{\citenamefont {Chen}\ \emph {et~al.}(2024)\citenamefont {Chen},
  \citenamefont {Rhee},\ and\ \citenamefont {Zwart}}]{2024-chen}%
  \BibitemOpen
  \bibfield  {author} {\bibinfo {author} {\bibfnamefont {B.}~\bibnamefont
  {Chen}}, \bibinfo {author} {\bibfnamefont {C.-H.}\ \bibnamefont {Rhee}},\
  and\ \bibinfo {author} {\bibfnamefont {B.}~\bibnamefont {Zwart}},\ }\bibfield
   {title} {\bibinfo {title} {Sample-path large deviations for a class of
  heavy-tailed markov-additive processes},\ }\href
  {https://doi.org/10.1214/24-EJP1115} {\bibfield  {journal} {\bibinfo
  {journal} {Electronic Journal of Probability}\ }\textbf {\bibinfo {volume}
  {29}},\ \bibinfo {pages} {1 } (\bibinfo {year} {2024})}\BibitemShut {NoStop}%
\bibitem [{\citenamefont {Bryson}(1974)}]{1974-exceedance}%
  \BibitemOpen
  \bibfield  {author} {\bibinfo {author} {\bibfnamefont {M.~C.}\ \bibnamefont
  {Bryson}},\ }\bibfield  {title} {\bibinfo {title} {Heavy-tailed
  distributions: properties and tests},\ }\href
  {https://doi.org/10.1080/00401706.1974.10489150} {\bibfield  {journal}
  {\bibinfo  {journal} {Technometrics}\ }\textbf {\bibinfo {volume} {16}},\
  \bibinfo {pages} {61} (\bibinfo {year} {1974})}\BibitemShut {NoStop}%
\bibitem [{\citenamefont {Naess}\ \emph {et~al.}(2013)\citenamefont {Naess},
  \citenamefont {Gaidai},\ and\ \citenamefont {Karpa}}]{2013-exceedance}%
  \BibitemOpen
  \bibfield  {author} {\bibinfo {author} {\bibfnamefont {A.}~\bibnamefont
  {Naess}}, \bibinfo {author} {\bibfnamefont {O.}~\bibnamefont {Gaidai}},\ and\
  \bibinfo {author} {\bibfnamefont {O.}~\bibnamefont {Karpa}},\ }\bibfield
  {title} {\bibinfo {title} {Estimation of extreme values by the average
  conditional exceedance rate method},\ }\href
  {https://doi.org/https://doi.org/10.1155/2013/797014} {\bibfield  {journal}
  {\bibinfo  {journal} {Journal of Probability and Statistics}\ }\textbf
  {\bibinfo {volume} {2013}},\ \bibinfo {pages} {797014} (\bibinfo {year}
  {2013})}\BibitemShut {NoStop}%
\bibitem [{\citenamefont {Kirschner}\ and\ \citenamefont
  {Gerhart}(1998)}]{Kirschner1998}%
  \BibitemOpen
  \bibfield  {author} {\bibinfo {author} {\bibfnamefont {M.}~\bibnamefont
  {Kirschner}}\ and\ \bibinfo {author} {\bibfnamefont {J.}~\bibnamefont
  {Gerhart}},\ }\bibfield  {title} {\bibinfo {title} {Evolvability},\ }\href
  {https://doi.org/10.1073/pnas.95.15.8420} {\bibfield  {journal} {\bibinfo
  {journal} {Proc. Natl. Acad. Sci. U. S. A.}\ }\textbf {\bibinfo {volume}
  {95}},\ \bibinfo {pages} {8420} (\bibinfo {year} {1998})}\BibitemShut
  {NoStop}%
\bibitem [{\citenamefont {Doyle}\ and\ \citenamefont
  {Csete}(2011)}]{Doyle2011}%
  \BibitemOpen
  \bibfield  {author} {\bibinfo {author} {\bibfnamefont {J.~C.}\ \bibnamefont
  {Doyle}}\ and\ \bibinfo {author} {\bibfnamefont {M.}~\bibnamefont {Csete}},\
  }\bibfield  {title} {\bibinfo {title} {Architecture, constraints, and
  behavior},\ }\href {https://doi.org/10.1073/pnas.1103557108} {\bibfield
  {journal} {\bibinfo  {journal} {Proc. Natl. Acad. Sci. U. S. A.}\ }\textbf
  {\bibinfo {volume} {108 Suppl 3}},\ \bibinfo {pages} {15624} (\bibinfo {year}
  {2011})}\BibitemShut {NoStop}%
\bibitem [{\citenamefont {Iyer-Biswas}\ \emph
  {et~al.}(2014{\natexlab{b}})\citenamefont {Iyer-Biswas}, \citenamefont
  {Crooks}, \citenamefont {Scherer},\ and\ \citenamefont {Dinner}}]{2014-PRL}%
  \BibitemOpen
  \bibfield  {author} {\bibinfo {author} {\bibfnamefont {S.}~\bibnamefont
  {Iyer-Biswas}}, \bibinfo {author} {\bibfnamefont {G.~E.}\ \bibnamefont
  {Crooks}}, \bibinfo {author} {\bibfnamefont {N.~F.}\ \bibnamefont
  {Scherer}},\ and\ \bibinfo {author} {\bibfnamefont {A.~R.}\ \bibnamefont
  {Dinner}},\ }\bibfield  {title} {\bibinfo {title} {Universality in stochastic
  exponential growth},\ }\href {https://doi.org/10.1103/PhysRevLett.113.028101}
  {\bibfield  {journal} {\bibinfo  {journal} {Phys. Rev. Lett.}\ }\textbf
  {\bibinfo {volume} {113}},\ \bibinfo {pages} {028101} (\bibinfo {year}
  {2014}{\natexlab{b}})}\BibitemShut {NoStop}%
\bibitem [{\citenamefont {Csete}\ and\ \citenamefont
  {Doyle}(2002)}]{2002-doyle}%
  \BibitemOpen
  \bibfield  {author} {\bibinfo {author} {\bibfnamefont {M.~E.}\ \bibnamefont
  {Csete}}\ and\ \bibinfo {author} {\bibfnamefont {J.~C.}\ \bibnamefont
  {Doyle}},\ }\bibfield  {title} {\bibinfo {title} {Reverse engineering of
  biological complexity},\ }\href {https://doi.org/10.1126/science.1069981}
  {\bibfield  {journal} {\bibinfo  {journal} {Science}\ }\textbf {\bibinfo
  {volume} {295}},\ \bibinfo {pages} {1664} (\bibinfo {year}
  {2002})}\BibitemShut {NoStop}%
\end{thebibliography}
\end{document}